\documentclass[prd,preprintnumbers,nofootinbib,onecolumn]{revtex4} 

\usepackage{graphicx}
\usepackage{dcolumn}
\usepackage{bm}
\usepackage{latexsym}
\usepackage{amsfonts}
\usepackage{amssymb}
\usepackage{amsmath}

 \newcommand{\bqq}{\begin{equation}}
 \newcommand{\eqq}{\end{equation}}
 \newcommand{\bqn}{\begin{eqnarray}}
 \newcommand{\eqn}{\end{eqnarray}}
 
 \newcommand{\lb}{\label}

\begin{document}

\preprint{IPMU11-0150}

\title{General relativity  limit of   Ho\v{r}ava-Lifshitz gravity  with a scalar
field in gradient expansion}

\author{A. Emir G\"umr\"uk\c{c}\"uo\u{g}lu}
\email{emir.gumrukcuoglu@ipmu.jp}
\affiliation{
 IPMU, The University of Tokyo, Kashiwa, Chiba 277-8582, Japan}
\author{Shinji Mukohyama}
\email{shinji.mukohyama@ipmu.jp}
\affiliation{
 IPMU, The University of Tokyo, Kashiwa, Chiba 277-8582, Japan}
\author{Anzhong Wang}
\email{Anzhong_Wang@baylor.edu}
\affiliation{
GCAP-CASPER, Physics Department, Baylor University, Waco, 
TX 76798-7316, USA}
\date{\today}

\begin{abstract}
 We present a fully nonlinear study of long wavelength cosmological
 perturbations within the framework of the projectable
 Ho\v{r}ava-Lifshitz gravity, coupled to a single scalar field. Adopting 
 the gradient expansion technique, we explicitly integrate the dynamical
 equations up to any order of the expansion, then restrict the
 integration constants by imposing the momentum constraint. While the
 gradient expansion relies on the long wavelength approximation,
 amplitudes of perturbations do not have to be small. When the 
 $\lambda\to 1$ limit is taken, the obtained nonlinear solutions exhibit
 a continuous behavior at any order of the gradient expansion,
 recovering general relativity in the presence of a scalar field and the
 ``dark matter as an integration constant''. This is in sharp contrast
 to the results in the literature based on the ``standard'' (and naive)
 perturbative approach where in the same limit, the perturbative
 expansion of the action breaks down and the scalar graviton mode
 appears to be strongly coupled. We carry out a detailed analysis on the 
 source of these apparent pathologies and determine that they originate
 from an improper application of the perturbative approximation in the
 momentum constraint. We also show that there is a new branch of
 solutions, valid in the regime where $|\lambda-1|$ is smaller than the
 order of perturbations. 
In the limit $\lambda\to1$, this new branch allows the theory to be continuously connected to general relativity, with an effective component which acts like pressureless fluid.
\end{abstract}

\maketitle

\section{Introduction}

Recently,  Ho\v{r}ava \cite{Horava} proposed a new theory of quantum
gravity in the framework of quantum field theory. One of the essential
ingredients of the theory is inclusion of higher-dimensional operators,
so that they dominate the ultraviolet (UV) behavior and render the
theory power counting renormalizable. Improvement of the UV behavior by
higher-dimensional operators has been known for some time \cite{Stelle}
but in those previous attempts, higher time derivative terms led to
ghost degrees of freedom. The major modification put forward by
Ho\v{r}ava's theory is that the power-counting renormalizability is
achieved without inclusion of higher time derivative terms. This is
realized by invoking the anisotropic scaling between time and space, 
\bqq
\lb{1.1}
t \rightarrow b^{-z} t,\;\;\; \vec{x} \rightarrow b^{-1}\vec{x}\,,
\eqq
so that higher-dimensional operators include spatial derivatives
only. This is reminiscent of Lifshitz scalars \cite{Lifshitz} in
condensed matter physics, hence the theory is often referred to as 
the Ho\v{r}ava-Lifshitz (HL) gravity. For the $3+1$ dimensional theory to be
power-counting renormalizable, the dynamical critical exponent $z$ has
to be larger than or equal to $3$ \cite{Horava} (see also
\cite{Visser}). Because of the anisotropic scaling, the theory cannot be
invariant under the spacetime diffeomorphism, 
${x}^{\mu} \rightarrow {x'}^{\mu}({x}^{\nu}),\; (\mu, \nu = 0, 1, 2, 3)$. 
Instead, the fundamental symmetry of the theory is the invariance under
the so called foliation-preserving diffeomorphism, 
\begin{equation}
\lb{1.2}
 t \to t'(t), \quad \vec{x}\to\vec{x}'(t,\vec{x})\,,
\end{equation}
denoted usually by Diff($M, \; {\cal{F}}$). The basic variables of the
theory are the lapse function $N$, the shift vector $N^i$, and the 
$3$-dimensional spatial metric $g_{ij}$  \cite{ADM}. Since the lapse
function $N$ corresponds to a gauge degree of freedom associated with
the space-independent time reparametrization, it is natural to restrict
the lapse function to be independent of the spatial coordinates: 
\bqq
\lb{1.4}
N = N(t)\,.
\eqq
This condition, imposed in the original formulation of the theory, is
called the {\em projectability condition}.

Since its introduction, there has been many cosmological applications of
the HL gravity and various remarkable features have been found (see
\cite{Mukohyama:2010xz,reviews} for reviews). In particular, the
higher-order spatial curvature terms can give rise to a bouncing
universe \cite{Calcagni}, may ameliorate the flatness problem \cite{KK}
and lead to caustic avoidance \cite{Mukohyama:2009tp}; the anisotropic
scaling provides a solution to the horizon problem and generation of
scale-invariant perturbations without inflation \cite{Mukohyama:2009gg},
a new mechanism for generation of primordial magnetic seed field
\cite{MMS}, and also a modification of the spectrum of gravitational
wave background via a peculiar scaling of radiation energy density
\cite{MNTY}; with the projectability condition, the lack of a local 
Hamiltonian constraint leads to ``dark matter as an integration
constant'' \cite{Mukohyama:2009mz}; in the parity-violating version of
the theory, circularly polarized gravitational waves can also be
generated in the early universe \cite{TS}.

Despite all of its remarkable features, the theory has been challenged
by significant questions. In particular, the Diff($M, \; {\cal{F}}$)
symmetry allows the existence of an additional spin-$0$ degree of
freedom, often called {\it scalar graviton}, and its fate is one of
important open issues. Actually, the scalar graviton is known to be
unstable either in the UV due to ghost instability or in the infrared
(IR) due to gradient instability~\cite{SVW,BPS, Ins}, depending on the
value of a coupling constant $\lambda$. In order to avoid the ghost
instability, $\lambda$ must satisfy either $\lambda<1/3$ or
$\lambda>1$. Precisely in these two ranges, the scalar graviton exhibits
gradient instability at long distances. We then have to tame this IR
instability by expansion of the universe~\cite{HWW,WW} or have to hide
it by the standard Jeans instability. One can formulate a condition
under which one of these happens~\cite{Mukohyama:2010xz}. Essentially,
the condition says that $\lambda$ must be sufficiently close to $1$ in
the IR.

However, in the limit $\lambda\to 1$, the scalar graviton appears to be
strongly coupled~\cite{SC,KP,WW}. That is, the ``standard'' (and naive) 
perturbative expansion breaks down in the sense that nonlinear terms
dominate linear terms in the $\lambda\to 1$ limit. Note that this
does not necessarily imply loss of predictability: if the theory is
renormalizable, all coefficients of infinite number of nonlinear terms
can be written in terms of finite parameters in the action, as several 
well-known theories with strong coupling (e.g. \cite{Pol}) indicate.
However, because of the breakdown of the (naive) perturbative expansion,
we need to employ nonperturbative methods to analyze the fate of the
scalar graviton in the limit  $\lambda \to 1$. Such an analysis was
performed in \cite{Mukohyama:2010xz} for spherically symmetric, static,
vacuum configurations and it was shown that the limit is continuously
connected to general relativity (GR). \footnote{Specifically, 
the solutions are continuously connected to the $\lambda=1$ theory,
whose action has the exact same form as the Einstein-Hilbert term (up to
high curvature terms negligible at low energies). However, due to the
different symmetries, the resulting theory is not exactly GR, but GR
with an effective component which acts like dark-matter
\cite{Mukohyama:2009mz}. This is what we mean by ``continuity with GR''
throughout the present paper. In the case considered in
\cite{Mukohyama:2010xz}, however, the ``dark matter'' component is
automatically set to zero by the assumed staticity.}
This may be considered as an analogue of the Vainshtein effect 
\cite{Vainshtein:1972sx,BDZ}. A similar consideration for cosmology was 
given in \cite{Izumi:2011eh}, where a fully nonlinear analysis of
superhorizon cosmological perturbations was carried out.

One of limitations of the analysis in \cite{Izumi:2011eh} is that it is
for a purely gravitational system in the absence of ordinary matter (but
with ``dark matter as an integration constant''). Since the naive 
perturbative expansion is known to break down not only in the gravity
sector but also in the matter sector \cite{SC,WW}, it is rather
important to extend the analysis of \cite{Izumi:2011eh} to the system
with ordinary matter. Technically speaking, however, this kind of
extention is indeed a nontrivial challenge since the system now has
multi components (ordinary matter and ``dark matter as an integration
constant'') and the gradient expansion technique has not been developed
for multi-component systems even in the standard cosmology in GR.

Thus, one of the main objectives of the present paper is to extend the
analysis of \cite{Izumi:2011eh} to the case where HL gravity is coupled
to a single scalar field, and provide yet another example indicating
that general relativity (plus ``dark matter as an integration
constant'') is restored in the $\lambda\to 1$ limit by nonlinear
dynamics. Another goal is to point out the source of the discrepancy
between perturbative and nonperturbative results. As we will see, the
solution of the momentum constraint in the naive application of the
``standard'' perturbative expansion is not valid in the regime where 
$|\lambda-1|$ is smaller than the order of perturbations.

The paper is organized as follows. In Sec.~\ref{basic}, we briefly
review the basic equations in the HL gravity with the projectability
condition (\ref{1.4}), while in Sec.~\ref{solutions}, we analyze the
inhomogeneous cosmology in HL gravity using the gradient expansion
method \cite{SBLMS}, and present the solutions to the equations of
motion. In Sec.\ref{SPA}, we present a discussion on the source of the
divergences in the naive perturbative expansion and show that the
momentum constraint is dominated by nonlinear terms in the $\lambda\to1$
limit. The results are summarized and discussed in
Sec.\ref{sec:summary}. The paper is supplemented by two Appendices, in
which we present some of the technical steps of our calculations.

\section{Basic equations}
\label{basic}

In this section, we review the basic equations of the HL gravity coupled
with a scalar field \cite{KK,WWM}, following the notation in
\cite{Mukohyama:2010xz}, and reformulate them in a way suitable for
gradient expansion~\cite{Izumi:2011eh}. In order to make the present
paper self-contained, some repetition of the material in
\cite{Izumi:2011eh} is inevitable in Secs.\ref{basic} and
\ref{solutions}, although we shall try our best to limit them to a
minimum. 

With Diff($M, \; {\cal{F}}$) and the projectability, the building blocks
of the theory are $g_{ij}$, $K_{ij}$,  $D_i$ and $R_{ij}$, where $K_{ij}$
denotes the extrinsic curvature of constant time hypersurfaces, $D_i$ is
the covariant derivative compatible with the the $3$-dimensional spatial
metric $g_{ij}$, and $R_{ij}$ is the three-dimensional Ricci tensor
built out of $g_{ij}$. (This is in contrast to GR or any other theory
with general covariance whose building blocks are the 4-dimensional
metric and its Riemann tensor.) For the critical exponent $z=3$, their
momentum dimensions are, respectively,  $[K_{ij}]=[k]^{3}$ and
$[R_{ij}]=[k]^{2}$. Throughout the present paper, we shall impose the
projectability condition as well as invariance under the spatial parity
($x^{i} \rightarrow - x^{i}$) and the time reflection  
($t \rightarrow - t$). The number of independent coupling constants in
this setup is $11$ for $z=3$~\cite{SVW,KK}. 
In fact, with the foliation-preserving diffeomorphisms (\ref{1.2}),  the
projectability condition (\ref{1.4}),  and the additional requirements of
parity and time reflection symmetry, 
 the most general gravitational
action can be specified as 
\begin{eqnarray}
I_g=\frac{M_{Pl}^2}{2}\int Ndt \, \sqrt{g}\,d^3\vec{x}
 \left(K_{ij} K^{ij} - \lambda K^2 -2\Lambda+R+L_{z>1}\right)\,,
\label{eq:action}
\end{eqnarray}
where $g$ is the determinant of $g_{ij}$, and  the
extrinsic curvature $K_{ij}$ is defined as
\begin{eqnarray}
K_{ij}=\frac{1}{2N}\left(
\partial_t g_{ij}-D_i N_j-D_j N_i
\right) \,,
\end{eqnarray}
$K$ ($=g^{ij}K_{ij}$) is the trace of $K_{ij}$, 
and $R$ is the Ricci
scalar constructed from $g_{ij}$. To lower and raise an index, $g_{ij}$
and its inverse $g^{ij}$ are used. For the sake of simplicity and
clarity, in the remainder of this paper, we  choose our units such
that $M_{Pl} = 1$.

In contrast to GR, the less restricting symmetry allows both kinetic
terms $K_{ij}K^{ij}$ and $K^2$  to be invariant independently, giving
rise to the extra parameter $\lambda$, which assumes the value $1$ in
GR, as mentioned above. Furthermore, in order to realize the
power-counting renormalizability, the higher curvature Lagrangian
$L_{z>1}$ should include up to sixth spatial derivatives. For the
analysis in the present paper, the concrete form of $L_{z>1}$ is not
needed. Adding the scalar field action $I_\phi$ that is invariant under
spatial parity and  time reflection, as well as the foliation preserving
diffeomorphism, the total action is 
\begin{eqnarray}
 I & = & I_g + I_{\phi}, \nonumber\\
 I_{\phi} & = & 
 \int N\,dt\,\sqrt{g}\,d^3\vec{x}
  \left[\frac{1}{2}(\partial_{\perp}\phi)^2-V(\phi,D_i,g_{ij})\right]\,,
\label{actphi}
\end{eqnarray}
where we define the derivative along vector normal to the hypersurface
\begin{equation}
 \partial_{\perp} \equiv \frac{1}{N}(\partial_t-N^k\partial_k)\,,
\end{equation}
and we decompose the scalar field potential as 
\begin{equation}
 V(\phi,D_i,g_{ij}) = V_0(\phi)
  +  V_{z\geq 1}(\phi,D_i,g_{ij})\,. 
\end{equation}
Here, $V_{z\geq 1}$ summarizes terms with two or more spatial
derivatives and like $L_{z>1}$ above, its concrete form is not needed for the purposes of the present paper.

Variation of the total action with respect to the 3-dimensional metric $g_{ij}$ leads to the dynamical equation
\begin{equation}
 {\cal E}_{g ij} + {\cal E}_{\phi ij} = 0\,,
\label{gijdynamical}
\end{equation}
where
\begin{eqnarray}
 {\cal E}_{g ij} & \equiv &
  g_{ik}g_{jl}\frac{2}{N\sqrt{g}}
  \frac{\delta I_g}{\delta g_{kl}}\nonumber\\
& = & 
  -\frac{1}{N}(\partial_t-N^kD_k)p_{ij}
  + \frac{1}{N}(p_{ik}D_jN^k+p_{jk}D_iN^k) 
  \nonumber\\
 & & 
  - Kp_{ij} + 2K_i^kp_{kj}
      + \frac{1}{2}g_{ij}K^{kl}p_{kl}-\Lambda \,g_{ij} 
      - G_{ij} 
 + {\cal E}_{g, z>1, ij}\,, \\
 {\cal E}_{\phi ij} & \equiv &
  g_{ik}g_{jl}\frac{2}{N\sqrt{g}}
  \frac{\delta I_{\phi}}{\delta g_{kl}}\nonumber\\
& = & 
g_{ij}\,\left[\frac{1}{2}\,\left(\partial_\perp\phi\right)^2 -V_0(\phi)\right] + {\cal E}_{\phi, z\geq 1, ij}\,. 
\end{eqnarray} 
Here, ${\cal E}_{g, z>1, ij}$ and ${\cal E}_{\phi, z\geq 1, ij}$ are 
contributions from $L_{z>1}$ and $-V_{z\geq 1}$, respectively, 
$p_{ij}\equiv K_{ij}-\lambda Kg_{ij}$, and $G_{ij}$ is Einstein tensor
of $g_{ij}$. The trace part and traceless part of Eq.(\ref{gijdynamical}) are,
respectively, 
\begin{equation}
 (3\,\lambda-1)\left(\partial_{\perp}K+\frac{1}{2}K^2\right)
  + \frac{3}{2}A^i_{\ j}A^j_{\ i} 
  + \frac{3}{2}(\partial_{\perp}\phi)^2 + Z = 0\,,
  \label{eq:Tr}
\end{equation}
and
\begin{equation}
 \partial_{\perp}A^i_{\ j} + KA^i_{\ j}
  + \frac{1}{N}(A^k_{\ j}\partial_kN^i-A^i_{\ k}\partial_jN^k)
  -\left(Z^i_{\ j}-\frac{1}{3}Z\delta^i_j\right) = 0\,,
\label{eq:Trless}
\end{equation}
where 
\begin{equation}
 A^i_{\ j} \equiv K^i_{\ j} - \frac{1}{3}K\delta^i_j\,,
\end{equation}
is the traceless part of $K^i_{\ j}$ and we defined
\begin{eqnarray}
 Z^i_{\ j} & \equiv & Z^i_{g,j} + Z^i_{\phi,j}\,, \qquad
  Z \equiv Z^i_{\ i}\,, \nonumber\\
 Z^i_{g,j} & \equiv & -\Lambda\,\delta^i_j - G^i_{\ j}
  + 
  g^{ik}{\cal E}_{z>1,g,kj}\,, \nonumber\\
 Z^i_{\phi,j} & \equiv & 
  -V_0(\phi)\delta^i_j
  + 
  g^{ik}{\cal E}_{z\geq 1,\phi,kj}\,.
\label{eqZdef}
\end{eqnarray}
Here, $Z^i_{g\,,j}$ is the variation of the potential part of the gravitational action with respect to the spatial metric; it is a generalization of (minus) the Einstein tensor of $g_{ij}$ to include higher curvature terms, as well as the cosmological constant. The quantity $Z^i_{\phi\,,j}$ is obtained similarly from the potential part of the scalar field action.

The variation of the total action with respect to $\phi$ yields the remaining dynamical equation
\begin{equation}
 0 = - \frac{1}{
 N\sqrt{g}}
  \frac{\delta I_{\phi}}{\delta\phi}
 = \frac{1}{N\sqrt{g}}\partial_t(\sqrt{g}\,\partial_{\perp}\phi)
  - \frac{1}{N\sqrt{g}}D_i(\sqrt{g}\,N^i\partial_{\perp}\phi)
  + E_{\phi}\,,\label{eqn:eomphi}
\end{equation}
where
\begin{equation}
 E_{\phi} \equiv
  \frac{1}{\sqrt{g}}\,\frac{\delta}{\delta\phi}
  \int\sqrt{g}\,dt\,d^3\vec{x}\ V(\phi,D_i,g_{ij})
  = V_0'(\phi) + E_{\phi,z\geq 1}\,,
\label{eqEdef}
\end{equation}
and $E_{\phi,z\geq 1}$ is the contribution from $V_{z\geq 1}$. 

Since the 3-dimensional spatial diffeomorphism is a subgroup of the foliation preserving diffeomorphism, $Z^i_{g,j}$ and
$Z^i_{\phi,j}$ satisfy the generalized Bianchi identity and matter conservation,
\begin{eqnarray}
D_j Z^j_{g,i}=0\,, \qquad
 D_j Z^j_{\phi,i} + E_{\phi}\partial_i\phi = 0 \,.
\label{eq:DZ}
\end{eqnarray}

For convenience, we decompose the spatial metric and the extrinsic curvature as
\begin{eqnarray}
g_{ij} & = & a^2(t) \, e^{2\zeta(t,\vec{x})} \,\gamma_{ij}(t,\vec{x})\,, 
\label{eq:decompositiongij}\\
K^i_{\ j} & = & \frac{1}{3} \,K(t,\vec{x}) \,\delta^i_{\ j}
 +A^i_{\ j}(t,\vec{x})\,,
 \label{eq:decomposition}
\end{eqnarray}
where we have defined $\zeta(t,\vec{x})$ so that $\det\gamma =1$, and
$a(t)$ (up to an overall normalization) is defined later in Eq.~(\ref{defa}).
The trace part and the traceless part of the definition of the
extrinsic curvature lead, respectively, to 
\begin{equation}
\partial_{\perp} \zeta+ \frac{\partial_t a}{N\,a}
 = \frac{1}{3}\, \left(K +\frac{1}{N}\partial_iN^i\right)\,,
\label{eq:TrK}
\end{equation}
and
\begin{equation}
 \partial_{\perp}\gamma_{ij}
  = 2\,\gamma_{ik}A^k_{\ j}
  + \frac{1}{N}
  \left(\gamma_{jk}\partial_iN^k
  + \gamma_{ik}\partial_jN^k
  -\frac{2}{3}\,\gamma_{ij}\partial_kN^k
       \right)\,.
\label{eq:TrlessK}
\end{equation}

The momentum constraint is obtained by varying the action with respect to $N^i$:
\begin{equation}
 D_jK^j_{\ i}- \lambda \,\partial_iK 
  = \partial_{\perp}\phi\,\partial_i\phi\,. 
\label{momcons}
\end{equation}
According to the decomposition (\ref{eq:decomposition}), the momentum
constraint is rewritten as
\begin{equation}
\partial_jA^j_{\ i} + 3 \, A^j_{\ i}\,\partial_j\zeta 
 - \frac{1}{2}A^j_{\ l} \, (\gamma^{-1})^{lk} \, \partial_{i}\gamma_{jk}
 - \frac{1}{3}\left(3\lambda-1\right)\partial_iK
 = \partial_{\perp}\phi\,\partial_i\phi\,.
\label{eq:momentum-constraint-pre}
\end{equation}

It can be shown that the evolution equations we have derived are
consistent with vanishing $A^i_{\ i}$, $\ln\det\gamma$,
$\gamma_{ij}-\gamma_{ji}$ and 
$\gamma_{ik}A^k_{\ j}-\gamma_{jk}A^k_{\ i}$~\cite{Izumi:2011eh}. 

\section{Gradient expansion}
\label{solutions}

In this section, we analyze the dynamics of nonlinear superhorizon
perturbations in the spatial gradient expansion approach. This approach
is valid as long as the characteristic length scale $L$ of the
perturbations is much larger than the Hubble length $H^{-1}$. By the
introduction of small parameter $\epsilon\sim 1/(H\,L)$, we perform a
series expansion on all relevant quantities and equations. For instance,
a spatial derivative acting on a quantity at order $\epsilon^p$ raises
the order to $\epsilon^{p+1}$ and thus is counted as ${\cal
O}(\epsilon)$. We then solve the equations order by order in gradient
expansion, extending the calculations of \cite{Izumi:2011eh} in a
spatially flat Friedmann-Robertson-Walker background, to include a
single scalar field as the source. 

\subsection{Gauge fixing}

The foliation preserving diffeomorphism invariance, like all other 
gauge symmetries, reflects a redundancy in the descriptions of the theory. 
By an appropriate choice of gauge conditions, these degrees can be eliminated 
and physical quantities can be extracted. In the present paper we adopt
the synchronous gauge, or the Gaussian normal coordinate system, by
setting the lapse function to unity and the shift vector to zero:
\begin{eqnarray}
N=1 \,,\qquad N^i= 0 \,.\label{eq:gaugefix}
\end{eqnarray}
This choice fixes the time coordinate but in the spatial coordinates,
there remains a gauge freedom of time-independent spatial
diffeomorphism, corresponding to the change of coordinates on the
initial constant-time hypersurface. This residual gauge degree of
freedom will be discussed later in Subsection~\ref{sec:physical}.

After the gauge fixing, our basic equations (\ref{eq:Tr}), (\ref{eq:Trless}),
(\ref{eqn:eomphi}), (\ref{eq:TrK}) and (\ref{eq:TrlessK}) are simplified
to
\begin{eqnarray}
(3\lambda-1) \partial_t K & = &
 -\frac{1}{2}(3\lambda-1)K^2 -\frac{3}{2}A^i_{\ j}A^j_{\ i}
 - \frac{3}{2}(\partial_t\phi)^2 - Z\,,
 \label{eq:K}\\
\partial_t A^i_{\ j} & = & 
 -KA^i_{\ j}+Z^i_{\ j}-\frac{1}{3}Z\,\delta^i_{\ j},\label{eq:A}\\
 0 & = & \partial_t^2\phi + K\partial_t\phi + E_{\phi}\,,\\
\partial_t \zeta & = & 
 -\frac{\partial_t a}{a}+\frac{1}{3}K \,,\label{eq:psi}\\
\partial_t \gamma_{ij} & = &
 2\,\gamma_{ik}A^k_{\ j}\,,\label{eq:evogama}
\end{eqnarray}
while the momentum constraint (\ref{eq:momentum-constraint-pre}) has 
the form 
\begin{equation}
\partial_j A^j_{\ i} + 3 \, A^j_{\ i}\partial_j\zeta 
 - \frac{1}{2}\,A^j_{\ l}(\gamma^{-1})^{lk}\partial_{i}\gamma_{jk}
 - \frac{1}{3}\,\left(3\lambda-1\right)\partial_iK
 = \partial_t\phi \,\partial_i\phi\,.
\label{eq:momentum-constraint}
\end{equation}
Hereafter, we assume that $\lambda\ne 1/3$; this is consistent with the
regime of physical interest $\lambda \ge 1$, discussed in the
Introduction section.

\subsection{Basic assumptions and order analysis}

We begin by determining the order of all relevant variables. In the
limit $\epsilon\to0$, we expect a universe that looks locally like a
Friedmann universe, leading to our starting assumption 
\begin{eqnarray}
 \partial_t \gamma_{ij} = {\cal O}(\epsilon) \,.\label{eqn:assumption1}
\end{eqnarray}
For the scalar field, a similar assumption leads to
$\partial_i\phi={\cal O}(\epsilon)$. However, in order to simplify the
analysis, we impose the stronger condition 
\begin{equation}
 \partial_i\phi = O(\epsilon^2). 
\end{equation}
That is, we assume that $\phi^{(0)}$, which is the leading order term of
$\phi$, is only time dependent: 
\begin{equation}
 \phi^{(0)} = \phi^{(0)}(t).
\end{equation}

The first assumption (\ref{eqn:assumption1}) then implies, from
Eq.~(\ref{eq:evogama}), 
\begin{eqnarray}
A^i_{\ j}=O(\epsilon),
\end{eqnarray}
leading, using the constraint equation (\ref{eq:momentum-constraint}), to
\begin{eqnarray}
\partial_i K=O(\epsilon^2).
\end{eqnarray}
In other words, the zero-th order part $K^{(0)}$ of $K$ depends on $t$
only. This fact enables us to define $a(t)$ by 
\begin{eqnarray}
3 \,\frac{\partial_t a(t)}{a(t)}= K^{(0)} (\equiv 3 H(t)).
\label{defa}
\end{eqnarray}
With this definition of $a(t)$, Eq.~(\ref{eq:psi}) leads to 
\begin{eqnarray}
 \partial_t \zeta=O(\epsilon). 
\end{eqnarray}

To summarize, the relevant quantities in the analysis are expanded as follows:
\begin{eqnarray}
 \zeta & = & \zeta^{(0)}(\vec{x}) +\epsilon \,\zeta^{(1)}(t,\vec{x})+ 
  \epsilon^2 \zeta^{(2)}(t,\vec{x})+
{\cal O}(\epsilon^3)
\,, \\
 \gamma_{ij} & = & f_{ij}(\vec{x}) 
  +\epsilon \,\gamma_{ij}^{(1)}(t,\vec{x})+
  \epsilon^2 \gamma_{ij}^{(2)}(t,\vec{x})+
{\cal O}(\epsilon^3)
 \,,\\
 K & = & 3 \,H(t)+\epsilon \,K^{(1)}(t,\vec{x})
  + \epsilon^2 K^{(2)}(t,\vec{x})+ 
{\cal O}(\epsilon^3)
\,, \\ 
 A^i_{\ j} & = & \epsilon \,A^{(1)\, i}_{\ \ \ \ \, j}(t,\vec{x}) +\epsilon^2
  A^{(2)\, i}_{\ \ \ \ \, j}(t,\vec{x}) + 
{\cal O}(\epsilon^3)
\,, \\
 \phi & = & \phi^{(0)}(t) + \epsilon\,\phi^{(1)}(t,\vec{x})
   + \epsilon^2\phi^{(2)}(t,\vec{x}) + 
{\cal O}(\epsilon^3)
\,,
\end{eqnarray}
where a quantity with the upper index $(n)$ corresponds to the $n$-th
order term in the gradient expansion.

\subsection{Equations in each order}

After determining the orders of all physical quantities, we now use this
information in the evolution equations (\ref{eq:K})--(\ref{eq:evogama})
to obtain the evolution equations at each order.  In the zero-th order
of gradient expansion we have 
\begin{eqnarray}
 (3\,\lambda-1)\left(\partial_t H + \frac{3}{2}\,H^2\right)
  & = & - \frac{1}{2}\,(\partial_t\phi^{(0)})^2
  + V_0(\phi^{(0)})+\Lambda \,, \nonumber\\ 
 \partial_t^2\phi^{(0)} + 3\,H \,\partial_t\phi^{(0)}
  + V'(\phi^{(0)}) & = & 0\,,
  \label{0thii}
\end{eqnarray}
where a prime denotes the ordinary derivative with respect to the indicated argument. %
By using the second of the above, the first equation can be integrated to give
\begin{equation}
 3\,H^2 = \frac{2}{3\,\lambda-1}
  \left[\frac{1}{2}\,(\partial_t\phi^{(0)})^2
   +V(\phi^{(0)})+\Lambda\right] + \frac{\widetilde{C}}{a^3},
\end{equation}
where $\widetilde{C}$ is an integration constant. The last term in the
right hand side of this equation is the ``dark matter as an integration
constant''~\cite{Mukohyama:2009mz}, a direct consequence of the
projectability condition. 

The dynamical equations at order ${\cal O}(\epsilon^n)$ with $n\geq 1$,
are written as 
\begin{eqnarray}
 a^{-3}\partial_t 
  \left[a^3 
   \left(K^{(n)}
    + \frac{3\,\phi^{(n)}\partial_t\phi^{(0)}}{3\,\lambda-1}
    \right)
   \right]
 & = & 
 -\frac{1}{2}\,\sum_{p=1}^{n-1} K^{(p)}K^{(n-p)} 
-\frac{3}{2\,(3\,\lambda-1)}
 \sum_{p=1}^{n-1}
 \left[
 A^{(p)\, i}_{\ \ \ \ \, j}A^{(n-p)\, j}_{\qquad\ \ i}
 \right.
 \nonumber\\
 & & 
  \left.
 +\partial_t\phi^{(p)}\partial_t\phi^{(n-p)}
 \right]
- \frac{\bar{Z}^{(n)}}{3\,\lambda-1}\,,
 \label{eq:n-K}\\
 a^{-3}\partial_t\left( a^3 A^{(n)\, i}_{\ \ \ \ \, j}\right) 
  & = &
  -\sum_{p=1}^{n-1}K^{(p)}A^{(n-p)\, i}_{\qquad\ \ j}
  +\bar{Z}^{(n)\, i}_{\ \ \ \ \, j}
  -\frac{1}{3}\,\bar{Z}^{(n)}\delta^i_{\ j}\,,
  \label{eq:n-A}\\
 a^{-3}\partial_t\left( a^3 \partial_t\phi^{(n)}\right) 
  +\left[V_0''(\phi^{(0)})
    -\frac{3\,(\partial_t\phi^{(0)})^2}{3\,\lambda-1}\right]
  \phi^{(n)}
  & = & 
  -\left(K^{(n)}+\frac{3\,\phi^{(n)}\partial_t\phi^{(0)}}{3\,\lambda-1}\right)
  \partial_t\phi^{(0)} \nonumber\\
 & & -\sum_{p=1}^{n-1}K^{(p)}\partial_t\phi^{(n-p)}
  -\bar{E}_{\phi}^{(n)}\,,
   \label{eq:n-phi}\\
 \partial_t \zeta^{(n)}
  & = & 
  \frac{1}{3} 
   \left(K^{(n)}
    + \frac{3\,\phi^{(n)}\partial_t\phi^{(0)}}{3\,\lambda-1}
    \right)
   - \frac{\phi^{(n)}\partial_t\phi^{(0)}}{3\,\lambda-1}\,,\\
 \partial_t \gamma^{(n)}_{ij}
  & = &
  2 \,\sum_{p=0}^{n-1}\gamma^{(p)}_{ik}
  A^{(n-p)\, k}_{\qquad\ \ j}\,,
  \label{eq:n-gamma}
\end{eqnarray}
where for later convenience, we introduced new (barred) quantities
\begin{equation}
 \bar{Z}^{(n)\, i}_{\quad\ \ j} \equiv 
  Z^{(n)\, i}_{\quad\ \ j}
  + V_0'(\phi^{(0)})\,\phi^{(n)}\delta^i_j, \quad
 \bar{Z}^{(n)} \equiv 
  Z^{(n)} + 3\,V_0'(\phi^{(0)})\,\phi^{(n)}, \quad
  \bar{E}_{\phi}^{(n)} \equiv 
  E_{\phi}^{(n)}-V_0''(\phi^{(0)})\,\phi^{(n)},
\label{eqbardef}
\end{equation}
to subtract the terms depending on $\phi^{(n)}$ from (unbarred) $Z^{(n)\,i}_{j}$ and $E^{(n)\,i}_j$, defined in Eqs.(\ref{eqZdef}) and (\ref{eqEdef}).
Here, $Z^{(n)\, i}_{\quad\ \ j}$, $Z^{(n)}$ and
$E_{\phi}^{(n)}$ are the $n$-th order terms of $Z^i_{\ j}$, $Z$,
$E_{\phi}$, respectively. With this definition, $\bar{Z}^{(n)\, i}_{\quad\ \ j}$,
$\bar{Z}^{(n)}$ and $\bar{E}_{\phi}^{(n)}$ do not depend on
$\zeta^{(n)}$, $\gamma_{ij}^{(n)}$, $K^{(n)}$, 
$A^{(n)\, i}_{\ \ \ \ \, j}$, nor on $\phi^{(n)}$.

Similarly, from Eq.~(\ref{eq:momentum-constraint})  we obtain the order ${\cal O}(\epsilon^{n+1})$ ($n\geq 1$) momentum constraint as 
\begin{eqnarray}
 \partial_j A^{(n)\, j}_{\ \ \ \ \ i}
  +3\, \sum_{p=1}^n A^{(p)\, j}_{\ \ \ \ \ i}\partial_j\zeta^{(n-p)}
  - \frac{1}{2}\,\sum_{p=1}^n\sum_{q=0}^{n-p}
  A^{(p)\, j}_{\ \ \ \ \ l}(\gamma^{-1})^{(q)\, lk}
  \partial_{i}\gamma^{(n-p-q)}_{jk} \nonumber\\
  -\frac{1}{3}\,(3\,\lambda-1)\,\partial_i 
   \left(K^{(n)}
    +\frac{3\,\phi^{(n)}\partial_t\phi^{(0)}}{3\,\lambda-1}\right)
  -\sum_{p=1}^{n-1}\partial_t\phi^{(p)}\partial_i\phi^{(n-p)}
  =0\,,
\label{eq:n-const}
\end{eqnarray}
where $(\gamma^{-1})^{(n)\, ij}$ is the $n$-th order term of the
inverse of $\gamma_{ij}$, i.e. the inverse $(\gamma^{-1})^{ij}$ is
expanded as 
\begin{equation}
(\gamma^{-1})^{ij} = 
 f^{ij}
 + \epsilon \,(\gamma^{-1})^{(1)\, ij}
 + \epsilon^2 (\gamma^{-1})^{(2)\, ij}
 + \ldots\,, 
\end{equation}
where $f^{ij}=(\gamma^{-1})^{(0)\, ij}$ is the inverse of $f_{ij}$. It
is straightforward to show that $(\gamma^{-1})^{(n)\, ij}$ ($n\geq 1$) satisfies
the following differential equation:
\begin{equation}
 \partial_t (\gamma^{-1})^{(n)\, ij} =
  -2\sum_{p=1}^n
  A^{(p)\, i}_{\quad\ \ k}
  (\gamma^{-1})^{(n-p)\, kj}\,. 
  \label{eqn:gamma-inv-nth-eq}
\end{equation}

In addition to the dynamical equations and momentum constraint, there
are also some useful identities. First, we expand the generalized
Bianchi identity (\ref{eq:DZ}) to obtain 
\begin{eqnarray}
 \partial_j \bar{Z}^{(n)\, j}_{\quad\ \ i}
  + 3\,\sum_{p=1}^n
  \left(\bar{Z}^{(p)\, j}_{\quad\ \ i}
   -\frac{1}{3}\,\bar{Z}^{(p)}\delta^j_{\ i}\right)\partial_j\zeta^{(n-p)}
  -\frac{1}{2}\sum_{p=1}^n\sum_{q=0}^{n-p}
  \bar{Z}^{(p)\, j}_{\quad\ \ l}(\gamma^{-1})^{(q)\, lk}
  \partial_i\gamma^{(n-p-q)}_{jk} \nonumber\\
 + \sum_{p=1}^{n-1}
  \left[\bar{E}_{\phi}^{(n-p)}+V_0''(\phi^{(0)})\,\phi^{(n-p)}\right]
  \partial_i\phi^{(p)} = 0\,, \label{eq:nth-DZ}
\end{eqnarray}
for $n\geq 1$. Next, expanding the conditions $A^i_{\ i}=0$,~ $\partial_i\ln\det\gamma=0$,~ $\gamma_{ij}-\gamma_{ji}=0$,~ 
$\gamma_{ik}A^k_{\ j}-\gamma_{jk}A^k_{\ i}=0$ and 
$A^i_{\ j}-\gamma_{jk}A^k_{\ l}(\gamma^{-1})^{li}=0$ 
leads to the following identities:
\begin{eqnarray}
& &
 A^{(n)\, i}_{\quad\ \ i} = 0\,, \qquad
  \sum_{p=0}^n(\gamma^{-1})^{(p)\, jk}
  \partial_i\gamma^{(n-p)}_{jk} = 0\,, \qquad
  \gamma^{(n)}_{ij}-\gamma^{(n)}_{ji}=0\,, \nonumber\\
 & & 
  \sum_{p=0}^{n-1}
  \left( \gamma^{(p)}_{ik}A^{(n-p)\, k}_{\qquad\ \ j}
   -\gamma^{(p)}_{jk}A^{(n-p)\, k}_{\qquad\ \ i}
       \right) = 0\,, \qquad
 A^{(n)\, i}_{\quad\ \ j} - 
  \sum_{p=0}^{n-1}  \sum_{q=0}^{n-p-1}
  \gamma^{(p)}_{jk}A^{(n-p-q)\, k}_{\qquad\quad\quad l}
  (\gamma^{-1})^{(q)\, li}
   = 0\,. 
 \label{eqn:nth-identities}
\end{eqnarray}

\subsection{${\cal O}(\epsilon)$ solution}
\label{subsec:1storder}

For ${\cal O}(\epsilon)$, Eqs.(\ref{eq:n-K})--(\ref{eq:n-gamma})  reduce to 
\begin{eqnarray}
\partial_t 
 \left[ a^3 \left( K^{(1)}
	     +\frac{3\,\phi^{(1)}\partial_t\phi^{(0)}}{3\,\lambda-1}
	    \right)\right] & = & 0\,,
\label{eq:leading-Tr-EoM} \\
\partial_t \left( a^3 A^{(1)\, i}_{\ \ \ \ \, j} \right) & = & 0\,,
\label{eq:leading-Trless-EoM}\\
a^{-3}\partial_t \left( a^3 \partial_t\phi^{(1)}\right) 
 +\left[V_0''(\phi^{(0)})
   -\frac{3\,(\partial_t\phi^{(0)})^2}{3\,\lambda-1}\right]\phi^{(1)}
 & = & - \left( K^{(1)}
	     +\frac{3\,\phi^{(1)}\partial_t\phi^{(0)}}{3\,\lambda-1}
	    \right)\partial_t\phi^{(0)}\,, 
 \label{eq:leading-phi-EoM}\\
\partial_t \zeta^{(1)} & = & 
 \frac{1}{3} 
 \left( K^{(1)}
  +\frac{3\,\phi^{(1)}\partial_t\phi^{(0)}}{3\,\lambda-1}
 \right) - \frac{\phi^{(1)}\partial_t\phi^{(0)}}{3\,\lambda-1}\,,
 \label{eq:leading-TrK}\\
\partial_t \gamma^{(1)}_{ij} & = &
 2 \,f_{ik} A^{(1)\, k}_{\ \ \ \ \, j}\,,
\label{eq:leading-TrlessK}
\end{eqnarray}
where from equations (\ref{eqZdef}), (\ref{eqEdef}) and (\ref{eqbardef}), we have at first order, 
$\bar{Z}^{(1)i}_{\quad\ j}=\bar{Z}^{(1)}=
\bar{E}_{\phi}^{(1)}=0$. 
Integrating the above equations, we obtain 
\begin{eqnarray}
 K^{(1)}+\frac{3\,\phi^{(1)}\partial_t\phi^{(0)}}{3\,\lambda-1}
  & = & \frac{C^{(1)}(\vec{x})}{a(t)^3}\,,
 \label{K1}\\
 A^{(1)\, i}_{\ \ \ \ \, j} & = & 
  \frac{C^{(1)\, i}_{\ \ \ \ \, j}(\vec{x})}{a(t)^3}\,,\\
 \phi^{(1)} & = & 
  \left[C^{(1)}(\vec{x})\int_{t_{\rm in}}^tdt'\,
   \frac{f_2(t')\,\partial_{t'}\phi^{(0)}(t')}{a(t')^3W(t')}
   +\phi_{\rm in}^{(1)}(\vec{x})
	       \right]f_1(t) \nonumber\\
 & & 
  + \left[-C^{(1)}(\vec{x})\int_{t_{\rm in}}^tdt'\,
   \frac{f_1(t')\,\partial_{t'}\phi^{(0)}(t')}{a(t')^3W(t')}
   +\dot{\phi}_{\rm in}^{(1)}(\vec{x})
	       \right]f_2(t)\,, \\
 \zeta^{(1)} & = & 
  \frac{C^{(1)}(\vec{x})}{3}
  \int^t_{t_{\rm in}} \frac{dt'}{a^3(t')}
  - \frac{1}{3\,\lambda-1}
  \int^t_{t_{\rm in}} dt'\ 
  \phi^{(1)}(t')\,\partial_{t'}\phi^{(0)}(t')
  + \zeta^{(1)}_{\rm in}(\vec{x}),\\
 \label{zeta1}
  \gamma^{(1)}_{ij} & = & 
  2 \,f_{ik}(\vec{x})\,C^{(1)\, k}_{\ \ \ \ \, j}(\vec{x})
  \int^t_{t_{\rm in}}\frac{dt'}{a^3(t')}
  + \gamma^{(1)}_{{\rm in}\, ij}(\vec{x})\,,
\label{gamma1}
\end{eqnarray}
where the integration ``constants'' 
$C^{(1)}$, $C^{(1)\, i}_{\ \ \ \ \, j}$,~ $\phi^{(1)}_{\rm in}$,~ $\dot{\phi}^{(1)}_{\rm in}$,~ $\zeta^{(1)}_{\rm in}$ and
$\gamma^{(1)}_{{\rm in}\, ij}$ depend only on the spatial coordinates $\vec{x}^i$
 and satisfy 
\begin{equation}
 C^{(1)\, i}_{\ \ \ \ \, i}=0\,,\qquad
 f_{ik}C^{(1)\, k}_{\ \ \ \ \, j} =  f_{jk}C^{(1)\, k}_{\ \ \ \ \, i}\,.
\end{equation}
The functions $f_i(t)$ ($i=1,2$) are two independent solutions of the 
homogeneous equation
\begin{equation}
 a^{-3}\partial_t(a^3\partial_t f_i) 
  + \left[V_0''(\phi^{(0)})
     -\frac{3\,(\partial_t\phi^{(0)})^2}{3\,\lambda-1}\right]f_i
  = 0\,; 
  \qquad f_1(t_{\rm in})=1\,,\quad f'_1(t_{\rm in})=0\,;
  \qquad f_2(t_{\rm in})=0\,,\quad f'_2(t_{\rm in})=1\,,
\end{equation}
and
\begin{equation}
 W(t) \equiv f_1(t)\,\partial_t f_2(t)-f_2(t)\,\partial_tf_1(t)\,.
\end{equation}
The two first order integration ``constants'', $\zeta^{(1)}_{\rm in}$
and $\gamma^{(1)}_{{\rm in}\, ij}$, can be absorbed into their zero-th
order counterparts, $\zeta^{(0)}_{\rm in}$ and 
$\gamma^{(0)}_{{\rm in}\, ij}$. Thus, without loss of generality, we can
set 
\begin{equation}
 \zeta^{(1)}_{\rm in} = 0\,, \qquad \gamma^{(1)}_{{\rm in}\, ij}=0\,. 
\label{firstconstants}
\end{equation}
Finally, the momentum constraint equation (\ref{eq:n-const}) with $n=1$
leads to the following relation among the remaining integration
constants, $C^{(1)}$,~ $C^{(1)\, i}_{\ \ \ \ \, j}$,~ $\zeta^{(0)}$ and
$f_{ij}$, 
\begin{equation}
\partial_j C^{(1)\, j}_{\ \ \ \ \, i}
 + 3\,C^{(1)\, j}_{\ \ \ \ \, i}\partial_j\zeta^{(0)}
 - \frac{1}{2}\,C^{(1)\, j}_{\ \ \ \ \, l}f^{lk}\partial_i f_{jk}
 - \frac{1}{3}\,\left(3\,\lambda-1\right)\partial_iC^{(1)} = 0.
\label{const-1}
\end{equation}
%
Note that
$\phi^{(1)}_{\rm in}(\vec{x})$ and $\dot{\phi}^{(1)}_{\rm in}(\vec{x})$ 
do not appear in this equation. The physical meaning of
$\phi^{(1)}_{\rm in}(\vec{x})$ and $\dot{\phi}^{(1)}_{\rm in}(\vec{x})$ 
are obvious:
\begin{equation} 
 \left.\phi^{(1)}\right|_{t=t_{\rm in}} = \phi^{(1)}_{\rm in}(\vec{x})\,, 
  \quad
 \left.\partial_t\phi^{(1)}\right|_{t=t_{\rm in}}
 = \dot{\phi}^{(1)}_{\rm in}(\vec{x})\,. 
\end{equation}

\subsection{${\cal O}(\epsilon^n)$ solution ($n\geq 1$)}

Equipped with the zero-th and first order solution, we can now determine
the general solutions at arbitrary order in gradient expansion. For any
$n \geq 1$, the solution to Eqs.(\ref{eq:n-K})--(\ref{eq:n-gamma}) is 
\begin{eqnarray}
 K^{(n)} + \frac{3\,\phi^{(n)}\partial_t\phi^{(0)}}{3\,\lambda-1}
 & = & 
 \frac{1}{a^3(t)}\int^t_{t_{\rm in}}dt' a^3(t')
 \left\{ - \frac{\bar{Z}^{(n)}(t',\vec{x})}{3\,\lambda-1}
  -\frac{1}{2}\,\sum_{p=1}^{n-1} K^{(p)}(t',\vec{x})K^{(n-p)}(t',\vec{x})
\right.\nonumber\\
 & & \left.  -\frac{3}{2\,(3\,\lambda-1)} 
  \sum_{p=1}^{n-1}
  \left[A^{(p)\, i}_{\ \ \ \ \, j}(t',\vec{x})\,
   A^{(n-p)\, j}_{\qquad\ \ i}(t',\vec{x})
   + \partial_{t'}\phi^{(p)}(t',\vec{x})
   \,\partial_{t'}\phi^{(n-p)}(t',\vec{x})
       \right]
  \right\}\,, 
 \label{eq:nthK} \\
 A^{(n)\, i}_{\ \ \ \ \, j}
  & = &
 \frac{1}{a^3(t)}\int^t_{t_{\rm in}}dt' a^3(t')
 \left[
  -\sum_{p=1}^{n-1}K^{(p)}(t',\vec{x})
  A^{(n-p)\, i}_{\qquad\ \ j}(t',\vec{x})
  +\bar{Z}^{(n)\, i}_{\ \ \ \ \, j}(t',\vec{x})
  -\frac{1}{3}\bar{Z}^{(n)}(t',\vec{x})\delta^i_{\ j}
  \right]\,, \label{eq:nthA} \\
 \phi^{(n)} & = & 
  f_1(t) \int_{t_{\rm in}}^tdt'
   \frac{f_2(t')r^{(n)}(t',\vec{x})}{W(t')}
  -f_2(t)\int_{t_{\rm in}}^tdt'
   \frac{f_1(t')r^{(n)}(t',\vec{x})}{W(t')}\,, \\
 \zeta^{(n)}
  & = & 
  \int^t_{t_{\rm in}}dt' 
  \left[\frac{1}{3}\,
   \left(
    K^{(n)}(t',\vec{x}) + 
    \frac{3\phi^{(n)}(t',\vec{x})\partial_{t'}\phi^{(0)}(t')}{3\,\lambda-1}
   \right)
   - \frac{\phi^{(n)}(t',\vec{x})\partial_{t'}\phi^{(0)}(t')}{3\,\lambda-1}
       \right]\,,\\
 \gamma^{(n)}_{ij}
  & = &
  2 \int^t_{t_{\rm in}}dt' \sum_{p=0}^{n-1}\gamma^{(p)}_{ik}(t',\vec{x})
  A^{(n-p)\, k}_{\qquad\ \ j}(t',\vec{x})\,, \label{eq:nthgamma}
\end{eqnarray}
where 
\begin{equation}
 r^{(n)}(t,\vec{x}) \equiv 
  \left(K^{(n)}(t,\vec{x})
   +\frac{3\,\phi^{(n)}(t,\vec{x})\partial_t\phi^{(0)}(t)}{3\,\lambda-1}\right)
  \partial_t\phi^{(0)}(t)
  + \sum_{p=1}^{n-1}K^{(p)}(t,\vec{x})\,\partial_t\phi^{(n-p)}(t,\vec{x})
  + \bar{E}_{\phi}^{(n)}(t,\vec{x})\,,
\end{equation}
and by redefining $C^{(1)}$,~ $C^{(1)\, i}_{\ \ \ \ \, j}$,~
$\phi^{(1)}_{\rm in}$,~ $\dot{\phi}^{(1)}_{\rm in}$,~ $\zeta^{(0)}$ and 
$f_{ij}$, we have set, respectively,
\begin{equation}
 \left. K^{(n)}\right|_{t=t_{\rm in}} =
 \left. A^{(n)\, i}_{\ \ \ \ \, j}\right|_{t=t_{\rm in}} = 
 \left.\phi^{(n)}\right|_{t=t_{\rm in}} = 
 \left.\partial_t\phi^{(n)}\right|_{t=t_{\rm in}} = 
 \left. \zeta^{(n)}\right|_{t=t_{\rm in}} =
 \left. \gamma^{(n)}_{ij}\right|_{t=t_{\rm in}} = 0\,.
\end{equation}
We remind that the first order constants have already been fixed in
Eq.(\ref{firstconstants}) by redefinition of $\zeta^{(0)}$ and $f_{ij}$,
respectively.

The initial condition for $\gamma^{(n)}_{ij}$ ($n\geq 1$) implies that 
$\left.\gamma_{ij}\right|_{t=t_{\rm in}}=f_{ij}$,~ 
$\left.(\gamma^{-1})^{ij}\right|_{t=t_{\rm in}}=f^{ij}$ and 
$\left.(\gamma^{-1})^{(n)\, ij}\right|_{t=t_{\rm in}}=0$ ($n\geq 1$). 
Therefore, for $n\geq 1$, the solution to Eq.(\ref{eqn:gamma-inv-nth-eq})
is 
\begin{equation}
 (\gamma^{-1})^{(n)\, ij} =
  -2\int_{t_{\rm in}}^t dt'\sum_{p=1}^n
  A^{(p)\, i}_{\quad\ \ k}
  (\gamma^{-1})^{(n-p)\, kj}\,. 
\end{equation}

As shown in Appendix~\ref{proof}, the solution
(\ref{eq:nthK})-(\ref{eq:nthgamma}) automatically satisfies the
($n+1$)-th order momentum constraint equation (\ref{eq:n-const}),
provided that the redefined integration constants ($C^{(1)}$, 
$C^{(1)\, i}_{\ \ \ \ \, j}$, $\zeta^{(0)}$, $f_{ij}$) satisfy
(\ref{const-1}) up to ${\cal O}(\epsilon^{n+1})$.

\subsection{Number of physical degrees of freedom}
\label{sec:physical}

The solution we obtained in the previous subsection involves a number of
functions depending only on spatial coordinates,
$\zeta^{(0)}(\vec{x})$,~ $f_{ij}(\vec{x})$, $C^{(1)}(\vec{x})$,~
$C^{(1)\, i}_{\ \ \ \ \, j}(\vec{x})$,~ $\phi^{(1)}_{\rm in}(\vec{x})$
and $\dot{\phi}^{(1)}_{\rm in}(\vec{x})$ which emerged as integration
``constants''. However, not all of the components are independent nor
physical. Firstly, they are subject to the constraint
(\ref{const-1}). Secondly, as stated just after Eq.
(\ref{eq:gaugefix}), our gauge condition (\ref{eq:gaugefix}) leaves
time-independent spatial diffeomorphism as a residual gauge
freedom. Therefore, the number of physical degrees of freedom included 
in each integration ``constant'' is 
%
\begin{eqnarray}
 \zeta^{(0)}(\vec{x}) & \ldots & 
  1 \mbox{ scalar growing mode }
  = 1 \mbox{ component }, \nonumber\\
 f_{ij}(\vec{x}) & \ldots & 
  2 \mbox{ tensor growing modes }
  = 5 \mbox{ components } - 3 \mbox{ gauge }, \nonumber\\
 C^{(1)}(\vec{x}) & \ldots & 
  1 \mbox{ scalar decaying mode }
  = 1 \mbox{ component }, \nonumber\\
 C^{(1)\, i}_{\ \ \ \ \, j}(\vec{x}) & \ldots & 
  2 \mbox{ tensor decaying modes }
  = 5 \mbox{ components } - 3 \mbox{ constraints }, \nonumber\\
 \phi^{(1)}_{\rm in}(\vec{x}), \dot{\phi}^{(1)}_{\rm in}(\vec{x}) & \ldots & 
  2 \mbox{ scalar modes }.
\end{eqnarray}
This is consistent with the fact that the HL gravity
includes not only a tensor graviton ($2$ propagating degrees of freedom)
but also a scalar graviton ($1$ propagating degree of freedom) and that
our system includes a scalar field ($1$ propagating degree of freedom)
as well.

\section{Perturbative vs nonperturbative approaches} 
\label{SPA}

In the previous section, we have derived solutions for nonlinear
perturbations in any order of gradient expansion. While gradient
expansion relies on the long wavelength approximation, amplitudes of
perturbations do not have to be small. Thus, our analysis in the
previous section is totally nonperturbative with respect to amplitudes
of perturbations. The dynamical equations and their solutions do not
suffer from any divergences in the $\lambda \to 1$ limit, and GR
 coupled with a scalar field and dark matter is safely
recovered in this limit.

This is in sharp contrast with results known in the literature based on
the ``standard'' (and naive) perturbative approach, in which pathologies 
such as divergences and strong coupling are found in the $\lambda\to 1$
limit. In this section, we shall see how this problem arises in the
``standard'' perturbative approach and why it becomes under control in our
nonperturbative approach. In the ``standard'' perturbative approach, all
relevant equations are expanded with respect to amplitudes of
perturbations, irrespective of sizes of coefficients in the
expansion. We shall see that, in the momentum constraint,
coefficients of terms linear in perturbations actually vanish in the
$\lambda\to 1$ limit and thus, for sufficiently small but nonvanishing
$|\lambda-1|$, the linear terms become less important than nonlinear
terms. Hence, neglecting nonlinear terms, blindly solving the linearized
momentum constraint and then taking the $\lambda\to 1$ limit
would be totally nonsense and lead to inconsistencies. This is precisely
the situation in the ``standard'' perturbative approach. Clearly, this is a
breakdown of the treatment based on the  ``standard'' perturbative
expansion but not of the HL theory itself. Indeed, as already stated
above, our nonperturbative analysis in the previous section does not
show any pathologies in the $\lambda\to 1$ limit. In the rest of this
section, we shall investigate these issues explicitly. For simplicity we
shall consider the cases without the scalar field (but with the built-in 
``dark matter as integration constant'').

\subsection{Breakdown of  standard perturbative expansion in the 
$\lambda\to 1$ limit} \label{subsec:cubicaction}

In this subsection let us briefly review the standard 
perturbative approach and see that, contrary to the nonperturbative
approach based on the gradient expansion in the previous section,  it
breaks down in the $\lambda\to 1$ limit.

Let us adopt the following metric ansatz in the transverse gauge, 
\begin{equation}
 N = 1, \quad N_i = \partial_i B + n_i, \quad
  g_{ij} = a^2e^{2\zeta_T}\left(e^h\right)_{ij}, 
  \label{eqn:transverse-gauge}
\end{equation}
where $n_i$ is transverse and $h_{ij}$ is transverse and traceless:
$\partial^in_i=0$, $\partial^ih_{ij}=0$ and $h^i_{\ i}=0$. Throughout 
this subsection, indices are raised and lowered by $\delta^{ij}$ and
$\delta_{ij}$. We introduce a small parameter $\bar{\epsilon}$,
consider $\zeta_T$, $B$, $n_i$ and $h_{ij}$ as quantities of
$O(\bar{\epsilon})$, and perform perturbative expansion with respect
to $\bar{\epsilon}$.

In the regime of validity of the  standard perturbative expansion, in order to
calculate the action up to cubic order, it suffices to solve the
momentum constraint up to the first order, which can be written in the form, 
\begin{equation}
 \partial_i\left[a^2(3\lambda-1)\partial_t\zeta_T-(\lambda-1)\triangle B\right]
  + \frac{1}{2}\triangle n_i = 0, 
\end{equation}
leading to 
\begin{equation}
 a^{-2}\triangle B = \frac{3\lambda-1}{\lambda-1}\partial_t\zeta_T,
  \quad n_i = 0, \label{eqn:constraint-perturbativesolution}
\end{equation}
where $\triangle \equiv\partial^i\partial_i$.

It is straightforward to calculate the kinetic action up to the third
order. The quadratic part $I_{kin}^{(2)}$ and the cubic part
$I_{kin}^{(3)}$ are~\cite{Gumrukcuoglu:2011xg}
\begin{eqnarray}
 I_{kin}^{(2)} & = & 
 \int dtd^3\vec{x} a^3
  \left( a^{-2}\partial_t\zeta_T\triangle B
		+\frac{1}{8}\partial_t h^{ij}\partial_t h_{ij}\right), 
  \nonumber\\
 I_{kin}^{(3)} & = & 
 \int dtd^3\vec{x} a^3
  \left[
   3\zeta_T\left(a^{-2}\partial_t\zeta_T\triangle B
		+\frac{1}{8}\partial_t h^{ij}\partial_t h_{ij}\right)
   + \frac{1}{2}a^{-4}\zeta_T\partial^i
   (\partial_iB\triangle B+3\partial^jB\partial_i\partial_jB)
   \right.
   \nonumber\\
 & & 
  \left.
  \qquad
   + \frac{1}{2}(a^{-2}\partial^kh^{ij}\partial_kB-3\partial_t h^{ij}\zeta_T)
   a^{-2}\partial_i\partial_jB
   -\frac{1}{4}a^{-2}\partial_t h^{ij}\partial_kh_{ij}\partial^kB
  \right]. \label{eqn:cubicaction}
\end{eqnarray}
When $B$ is eliminated by using
(\ref{eqn:constraint-perturbativesolution}), one can easily see that the
quadratic part $I_{kin}^{(2)}$ written in terms of 
$\tilde{\zeta}_T=\sqrt{\frac{2(3\lambda-1)}{\lambda-1}}\zeta_T$ 
is regular. On the other hand, the cubic part $I_{kin}^{(3)}$ written in
terms of $\tilde{\zeta}_T$ is divergent in the limit $\lambda\to 1$. Thus,
the perturbative expansion breaks down in this limit. More precisely,
the regime of validity of the  standard perturbative expansion is 
\begin{equation}
 |\zeta_T| \ll \min(|\lambda-1|,1), \label{eqn:validity-perturbation}
\end{equation}
and disappears in the $\lambda\to 1$ limit.

Evidently, the breakdown of the  standard perturbative expansion in the
$\lambda\to 1$ limit originates from the denominator $\lambda-1$ in the solution
(\ref{eqn:constraint-perturbativesolution}) to the linearized momentum
constraint.

\subsection{Transformation from transverse to synchronous gauge}

In the  standard perturbative approach summarized in the previous
subsection, we have adopted the transverse gauge
(\ref{eqn:transverse-gauge}). Instead, in the nonperturbative approach
based on the gradient expansion presented in Sec.\ref{solutions}, 
we have adopted the synchronous gauge
(\ref{eq:gaugefix}). In this subsection, we shall investigate the
spatial coordinate transformation between the two gauges. (Note that in
both gauges the space-independent time reparametrization is already
fixed by the condition $N=1$.) The transformation is nonlinear but we
treat it perturbatively. As we shall see below, this provides an
alternative way to see the breakdown of the  standard perturbative
expansion.

As described in Appendix \ref{app:expansion}, we start with the
transverse gauge, carry out the spatial gauge transformation to the
synchronous gauge, and use the momentum constraint  (in the
transverse gauge) to eliminate the nondynamical degree of freedom. In
this way, we can express the perturbation in the synchronous gauge in
terms of that in the transverse gauge. Up to the second order, the
result is 
\begin{eqnarray}
\zeta &=& 
 -\frac{2}{3}\,\frac{1}{\lambda-1}
 \Bigg\{\zeta_T 
 - \frac{\left(3\,\lambda-1\right)}{(\lambda-1)}\,
 (\partial_i\zeta_T)\,(\partial^i \triangle^{-1}\zeta_T) 
+\frac{\left(3\,\lambda-1\right)}{(\lambda-1)}\,\, \int ^t dt'\,\left(\partial_i \zeta_T\right) \left(\partial^i\,\triangle^{-1}\partial_{t'}\zeta_T\right)
\nonumber\\
&&\qquad\quad\;\, -\frac{\left(3\,\lambda-1\right)}{2\,(\lambda-1)}
\int^t dt'\triangle^{-1}\left[2\,(\partial^i \triangle\zeta_T)(\partial_i \triangle^{-1}\partial_{t'}\zeta_T) + \left(\partial^i\partial^j \zeta_T+ \frac{1}{2}\,\triangle h^{ij}\right) (\partial_i \partial_j \triangle^{-1}\partial_{t'}\zeta_T)    + (\triangle\zeta_T)(\partial_{t'} \zeta_T) \right] \nonumber\\
&&\qquad\quad\;\,+\frac{1}{4}\,\int^t dt'\,\triangle^{-1}\left[ \frac{1}{2}\,(\partial_i \partial_{t'} h_{jk})(\partial^i h_{jk}) + \frac{1}{2}\,(\partial_{t'} h^{ij})(\triangle h_{ij}) -3\,(\partial_i\partial_j\zeta_T)(\partial_{t'} h^{ij})\right]
+ {\cal O}(\bar{\epsilon}^3)\Bigg\}\,,
\label{zetaNL}
\end{eqnarray}
where $\zeta$ is the perturbation in the synchronous gauge defined in
(\ref{eq:decompositiongij}), $\zeta_T$ and $h_{ij}$ are the metric
perturbations in the transverse gauge defined in
(\ref{eqn:transverse-gauge}). One can easily see that the terms
quadratic in $\zeta_T$ are suppressed with respect to the linear term
under the condition (\ref{eqn:validity-perturbation}).

Conversely, for a fixed amplitude of $\zeta_T$, the expansion with
respect to $\bar{\epsilon}$ in (\ref{zetaNL}) breaks down in the
$\lambda\to 1$ limit. This is very similar to the way how the  standard
perturbative expansion of the action (\ref{eqn:cubicaction}) breaks down
in the $\lambda\to 1$ limit. It is apparent from the intermediate steps
of the calculation (shown explicitly in Appendix~\ref{app:expansion})
that the terms with negative powers of $(\lambda-1)$ are introduced by
the solution of the momentum constraint.

\subsection{Linear vs nonlinear terms in the momentum constraint}
\label{subsec:linvsnonlin}

Having understood that the origin of the breakdown of the  standard
perturbative expansion is the treatment of the momentum constraint, 
we now discuss the regime of validity of the  standard perturbative expansion in
the momentum constraint. Importantly, we shall see that a new 
branch of solution emerges at the edge of the regime of validity of the
 standard perturbative expansion.

For this purpose, we adopt the transverse gauge
(\ref{eqn:transverse-gauge}) and expand the momentum constraint
with respect to $\zeta_T$ and $h_{ij}$,  considering them as small
quantities but keeping $B$ and $n_i$ as nonlinear quantities
\footnote{
Although the discussion in this subsection employs the transverse gauge,
the general argument holds in any gauge. In particular, in the
synchronous gauge, where $B=n_i=0$, the issue arising in $B$ gets
transferred to the longitudinal part of $h_{ij}$. However, for the sake
of clarity, we chose to keep find solutions for the nondynamical fields,
gauging away the longitudinal mode instead. 
}: 
\begin{equation}
 \zeta_T = O(q), \quad h_{ij} = O(q), \quad B=O(q^0), \quad n_i=O(q^0),
\end{equation}
where we have introduced a small parameter $q$ to count the order of
perturbations $\zeta_T$ and $h_{ij}$. In the absence of the scalar field,
the momentum constraint is \cite{Izumi:2011eh}
\begin{equation}
 0 = {\cal H}_j \equiv \partial_jA^j_{\ i} + 3 \, A^j_{\ i}\,\partial_j\zeta_T 
 - \frac{1}{2}A^j_{\ l} \, (\gamma^{-1})^{lk} \, \partial_{i}\gamma_{jk}
 - \frac{1}{3}\left(3\lambda-1\right)\partial_iK,  
\end{equation}
where $\gamma_{ij}=(e^{h})_{ij}$, $(\gamma^{-1})^{ij}=(e^{-h})^{ij}$, while the trace of the extrinsic curvature (\ref{eq:TrK}) becomes
\begin{eqnarray}
 K & = & 
  3\left(\partial_{\perp} \zeta_T+ \frac{\partial_t a}{Na}\right)
  - \frac{1}{N}\partial_i\left[(g^{-1})^{ij}N_j\right] \nonumber\\
 & = & 3H - a^{-2}\triangle B
  +3\partial_t\zeta_T
  + a^{-2}
  \left[-(\partial^k\zeta_T)(\partial_kB)
   + 2\zeta_T\triangle B + h^{kl}\partial_k\partial_lB
       \right] \nonumber\\
 & & + O(q)\times n_k + O(q^2),
\end{eqnarray}
and the traceless part (\ref{eq:TrlessK}) is
\begin{eqnarray}
 A^j_{\ i} & = & 
  \frac{1}{2}(\gamma^{-1})^{jk}\partial_{\perp}\gamma_{ki}
  - \frac{1}{2N}
  \left\{(\gamma^{-1})^{jk}\gamma_{il}\partial_k
   \left[(g^{-1})^{lm}N_m\right]
   + \partial_i
   \left[(g^{-1})^{jk}N_k\right]
   -\frac{2}{3}\delta^j_i\partial_k
   \left[(g^{-1})^{kl}N_l\right]
  \right\} \nonumber\\
 & = & 
  \frac{1}{2}\partial_t h^j_{\ i}  
  -\frac{1}{a^2}
  \left\{\frac{1}{2}(\partial^jn_i+\partial_in^j)
   + \left(\partial^j\partial_iB
      -\frac{1}{3}\delta^j_i\triangle B\right)\right\}
  \nonumber\\
& & 
 + \frac{1}{a^2}
  \left\{
   (\partial^j\zeta_T)(\partial_iB)+(\partial_i\zeta_T)(\partial^jB)
    +2\zeta_T(\partial^j\partial_iB)
   -\frac{2}{3}\delta^j_i
   \left[(\partial^k\zeta_T)(\partial_kB)+\zeta_T\triangle B\right]
	\right\}\nonumber\\
 & & 
 + \frac{1}{a^2}
  \left\{
      \frac{1}{2}
      \left(\partial_ih^{jk}
       + \partial^jh_i^{\ k}
       -\partial^kh^j_{\ i}\right)
      (\partial_kB)
      + h^{jk}(\partial_i\partial_kB)
      - \frac{1}{3}\delta^j_ih^{kl}(\partial_k\partial_l B)
		 \right\} \nonumber\\
 & & + O(q)\times n_k + O(q^2). 
\end{eqnarray}
Here, it is understood that $(g^{-1})^{ij}$ is the inverse of $g_{ij}$,
that derivatives do not act beyond parentheses and that indices are
raised and lowered by $\delta^{ij}$ and $\delta_{ij}$. A straightforward 
calculation results in the following expansion of ${\cal H}_i$,
\begin{eqnarray}
{\cal H}_i &=& 
 -\left(3\,\lambda-1\right) \partial_i \partial_t\zeta_T + {\cal O}(q^2)
 - \frac{1}{2\,a^2}\,
 \left[\triangle  +{\cal O}(q)\right] n_i
 \nonumber\\
 && + \frac{1}{a^2} 
  \left\{ 
   \left(\lambda-1\right)
   \left[\delta^j_i \triangle  +{\cal O}(q)\right]
   + \left(\frac{1}{2}\,\triangle  h^j_{\ i}
      + \partial^j\partial_i \zeta_T
      + \delta^j_i\,\triangle \zeta_T\right)
   + {\cal O}(q^2)\right\}\partial_jB \,.
\label{expandedcons}
\end{eqnarray}
Notice that in the above, no assumption is made for $B$ and $n_i$, which
are still considered to be nonlinear quantities. It is now clear that
the leading term in the coefficient of $B$ relies not only on the order of perturbations, but also on the value of $\lambda-1$.

In the regime $q\ll{\rm min}\left(1,\,\lambda-1\right)$, the momentum
constraint becomes 
\begin{equation}
a^{-2}\triangle  B = \,\frac{3\lambda-1}{\lambda-1}\,\partial_t \zeta_T
 + {\cal O}(q^2) \,,\qquad n_i = {\cal O}(q^2)\,,\qquad {\rm
 for~}q\ll{\rm min}\left(1,\,\lambda-1\right)\,,
 \label{eqn:solB-linear}
\end{equation}
and agrees with the result of the  standard perturbative expansion
(\ref{eqn:constraint-perturbativesolution}). Naively using this
expression in the action, then taking the $\lambda\to 1$ limit would
lead to breakdown of the  standard perturbative expansion as already seen in 
(\ref{eqn:cubicaction}) and (\ref{zetaNL}).

On the other hand, if $\lambda$ is sufficiently close to $1$ and the
condition 
\begin{equation}
\lambda-1 \ll q \ll 1 \label{eqn:nonlinear-regime}
\end{equation}
is met, then the coefficient of $B$ in Eq.(\ref{expandedcons}) is
dominated by the ${\cal O}(q)$ terms instead of the 
${\cal O}(\lambda-1)$ term. Note that this is a nonlinear regime but
is still consistent with the assumed smallness of the metric
perturbations $\zeta_T$ and $h_{ij}$. In this regime, the constraint can
be written as 
\begin{eqnarray}
{\cal H}_j &=& -\,2\partial_j\partial_t\zeta_T + {\cal O}(q^2) 
- \frac{1}{2\,a^2}\,\left(\triangle  +{\cal O}(q)\right) n_j
+ \frac{1}{a^2}\left[M^{\;\;i}_j + {\cal O}(\lambda-1) 
		+{\cal O}(q^2)\right]\partial_i B\,,
\label{IRconstraint}
\end{eqnarray}
where we have defined
\begin{equation}
M_j^{\;\;i} \equiv \frac{1}{2}\,\triangle  h^i_{\;j} +
 \partial^i\partial_j \zeta_T+ \delta^i_{j}\,\triangle  \zeta_T =
 O(q)\,.
\end{equation}
The transverse part of the Eq.(\ref{IRconstraint}) can be found by
evaluating $\partial_{[k}{\cal H}_{j]}$, 
\begin{eqnarray}
\partial_{[k} \triangle n_{j]} & = &
 (\triangle h^i_{\;\;[j})(\partial_{k]}\partial_i B)
  + (\partial_{[k} \triangle h^i_{\;\;j]})(\partial_i B)
  \nonumber\\
& & 
 + 2(\partial^i\partial_{[j}\zeta_T)(\partial_{k]}\partial_i B)
       +2(\partial_{[k}\triangle \zeta_T)(\partial_{j]}B)
       +{\cal O}(q^2).
\end{eqnarray}
On the other hand, the longitudinal part can be computed from
$\partial^j {\cal H}_j$ as 
\begin{equation}
-2\,\triangle\,\partial_t\zeta_T +\frac{1}{a^2}\,\bar{M}\,B=0\,,
\end{equation}
where we define the operator $\bar{M}$ as, 
\begin{equation}
\bar{M}\equiv 
 M^{ij}\partial_i\partial_j
 + 2\,(\partial^i\triangle\zeta_T)\,\partial_i\,.
\end{equation}
If this operator is invertible,  then we obtain
\begin{equation}
B = 2\,a^2\, \bar{M}^{-1}\,\triangle\partial_t\zeta_T 
 +{\cal O}\left(\frac{\lambda-1}{q}\right)+{\cal O}(q)\,.
 \label{eqn:solB-nonlinear}
\end{equation}
We note that either the $(\lambda-1)/q$ or $q$ term can provide the
largest correction to $B$, depending on the value of $\lambda-1$. 

In summary, we have seen that there are two branches of
solutions to the momentum constraint, depending on the value of 
$\lambda-1$. One is (\ref{eqn:solB-linear}) in the linear regime, and
the other is (\ref{eqn:solB-nonlinear}) in the nonlinear regime. The
standard  perturbative expansion in the previous subsections corresponds
to the  solution  (\ref{eqn:solB-linear}). On the other hand, what is
relevant for the nonperturbative recovery of GR (plus ``dark matter'')
in the $\lambda\to 1$ limit is the solution
(\ref{eqn:solB-nonlinear}). The two regimes are mutually exclusive.

\subsection{Yet another consideration}

In the previous subsection we have seen that there are two mutually
exclusive branches of solution to the momentum constraint. This explains
the reason why the  standard  perturbative approach breaks down in the
$\lambda\to 1$ limit and why the theory itself can be still regular and
continuous in the limit.

For $\lambda$ away from $1$, the standard perturbative expansion is
valid in the transverse gauge and we have the expansion of the kinetic
action as given in subsection \ref{subsec:cubicaction}. On the other
hand, for $\lambda$ sufficiently close to $1$, i.e. in the regime
(\ref{eqn:nonlinear-regime}), it is the nonlinear solution
(\ref{eqn:solB-nonlinear}) that should be substituted to the kinetic
action.

Unlike the kinetic action, the potential part of the action does not
depend on $\lambda$, when written in terms of $\zeta_T$ and 
$h_{ij}$. This is because $B$ does not appear in the potential part of
the action. Therefore, if we could somehow define a field $\zeta_c$ in
such a way that the series of terms in the kinetic action for $\zeta_T$
sums up to form a standard canonical kinetic term for $\zeta_c$, then
the whole action written in terms of $\zeta_c$ should remain finite in
the $\lambda\to 1$ limit. Since each term in the kinetic action (after
eliminating $B$) includes exactly two time derivatives, such a field
redefinition should be possible in principle. In practice, however, the
field redefinition is not easy to perform since it would be nonlinear
and highly nonlocal in space. Nonetheless, this consideration already
suggests that the $\lambda\to 1$ limit should be regular and continuous 
nonperturbatively.

\section{Summary and Discussion}
\label{sec:summary}

In this paper, we have performed a fully nonlinear analysis of
superhorizon perturbations in the HL gravity coupled to a scalar field,
by using the gradient expansion technique \cite{SBLMS}. After applying
the long wavelength expansion to the set of field equations, we
integrated these explicitly to the second order. We then showed that the
solutions can be extended to any order in gradient expansion, while
satisfying the momentum constraint at each order. These solutions are
continuous in the GR limit $\lambda \to 1$ for any order in the
expansion, both in the gravity sector, which consists of the ``dark
matter as an integration constant'', and in the matter sector,
which contains a scalar field in the present work. The form of the
equations suggests that our qualitative result should remain the same
when additional matter fields are introduced.

This is in sharp contrast with the results obtained in the framework of
the ``standard'' (and naive) perturbation theory, in which pathologies
such as divergences and strong coupling are found in the $\lambda\to 1$
limit \cite{SC,WW}. We determined that the results of the standard
perturbative expansion are valid only in the region where $|\lambda-1|$
is larger than the order of perturbations. In other words, the range of
validity (\ref{eqn:validity-perturbation}) of these solutions has zero
measure in the limit $\lambda\to1$. We found that the divergences are 
originating from the momentum constraint, where the coefficients of the
terms linear in perturbations vanish in the $\lambda\to1$ limit. Thus,
for sufficiently small but nonvanishing $|\lambda-1|$, the linear terms
become less important in comparison to the nonlinear ones. Neglecting
nonlinear terms and naively solving the linearized momentum constraint,
then taking the $\lambda\to 1$ limit turns out to be the main source of
the said pathologies. Once their origins were understood, we carried out
a detailed analysis of the nonlinear momentum constraint in the
perturbative approach. In addition to the known result which is valid
when $|\zeta| \ll \min(|\lambda-1|,1)$, we found a second branch of
solution valid in the regime where $|\lambda-1| \ll |\zeta| \ll 1)$. The
presence of the latter solution justifies the recovery of GR obtained in
our nonperturbative approach, in the $\lambda\to 1$ limit.

Our results, together with the similar examples studied in
\cite{Mukohyama:2010xz,Izumi:2011eh}, discernibly support that the 
apparent strong coupling found previously in the HL gravity may only
indicate the breakdown of the treatment based on the naive perturbative
expansion but not of the theory itself. General relativity should be
recovered by nonlinear effects, analogously to the Vainshtein mechanism
\cite{Vainshtein:1972sx} that were first encountered in the massive
gravity theories.

We note that the present analysis was limited to the discussion of the
classical (and superhorizon) evolution of perturbations; their quantum 
mechanical origin were not considered. In contrast, in
Ref.\cite{Gumrukcuoglu:2011xg}, the HL gravity in the $\lambda\to\infty$ 
limit was found to be weakly coupled under a certain condition, and the
spectrum of perturbations that were generated from quantum fluctuations
was calculated in this limit. However, the presence of regular behavior
both in $\lambda\to \infty$ and in $\lambda\to1$ limits is not
sufficient to draw conclusions on the transition between the two
regimes. This is because of our lack of an understanding on the details
of the renormalization group (RG) flow. Specifically, to be able to
match these two results, one needs to define a conserved quantity (like
the comoving curvature perturbations in relativistic
cosmology). However, since the matching involves a wide range of varying
$\lambda$, one needs to know how the flow of $\lambda$ is realized and
how such a flow affects the evolution of cosmological
perturbations. With these considerations, we refrain from exploring the
cosmological implications of our results for now.

On the other hand, a quantum mechanical extension of our analysis may
have a chance to address such issues. One of the major concerns with a
proper renormalization analysis in HL gravity is the strong coupling
problem in the $\lambda\to1$ limit, or more specifically, the breakdown
of the perturbative expansion. However, we have shown in
Sec.~\ref{subsec:linvsnonlin} that the full nonlinear analysis in the
limit $\lambda\to 1$ is still consistent with small perturbations,
except for the nondynamical mode $B$ which becomes nonlinear. The
solution to the momentum constraint in the two regimes, 
(\ref{eqn:solB-linear}) and (\ref{eqn:solB-nonlinear}), gives the
nondynamical mode as 
\begin{equation}
B \simeq \Bigg\{
\begin{array}{lll}
 \frac{3\lambda-1}{\lambda-1}\,a^2\triangle^{-1}\partial_t \zeta
  = O(\zeta)\,, 
  & \quad{\rm for} & |\zeta| \ll \min(\lambda-1,\,1)\\\\
 2\,a^2\, \bar{M}^{-1}\,\triangle\partial_t\zeta = O(1)\,, 
  & \quad{\rm for} & \lambda-1 \ll |\zeta| \ll 1
\end{array}\,. \label{eqn:solB}
\end{equation}
Note that both cases are compatible with small $\zeta$. Thus,
substituting this nonlinear solution for $B$ in the action, then
applying the perturbative expansion for $\zeta$ may provide a healthy
perturbative action. (However, the reduced action is nonlocal in space
while it is local in time).

We also note that the breakdown of the naive perturbative expansion
does not necessarily result in loss of renormalizability. We know that
in the regime $|\zeta|\ll\min(\lambda-1,\,1)$, the leading UV
contributions in the action are invariant under the scaling (\ref{1.1})
with $z=3$, provided that the scalar graviton is assigned a vanishing
scaling dimension, i.e. $\zeta\to\zeta$. This fact is nothing but the 
power-counting renormalizability of the theory, and is seen after
replacing $B$ in the action with the linear solution
(\ref{eqn:solB-linear}), or the first line of (\ref{eqn:solB}). Note
also that coefficients of all possible terms in the perturbative
expansion are expressed in terms of $11$ coupling constants in the
action (\ref{eq:action}). On the other hand, for the regime
$\lambda-1\ll|\zeta|\ll 1$, we need to replace $B$ in the action with
the nonlinear solution (\ref{eqn:solB-nonlinear}), or the second line of
(\ref{eqn:solB}). What is important here is that the scaling dimension
of $B$ from the nonlinear solution and that from the linear solution are
exactly the same: $B\to b\,B$ under the scaling (\ref{1.1}) with $z=3$
in both cases. Therefore, after substituting the nonlinear solution to
$B$ in the action, we still conclude that the leading UV contributions
in the action are invariant under the scaling (\ref{1.1}) with $z=3$,
provided that the scalar graviton is assigned a vanishing scaling
dimension. In other words, the conditions for power-counting
renormalizability of the theory continue to hold in the nonlinear
regime.

In the present paper, we have considered the projectable version of the
HL theory and showed that the general relativity (plus dark matter) is
safely recovered in the $\lambda\to 1$ limit. If we relax the
projectability condition and thus allow the lapse function to depend on
spatial coordinates then we should include as the building blocks of the
theory not only $g_{ij}$, $K_{ij}$, $D_i$ and $R_{ij}$ but also
$a_{i}\equiv\partial_{i}\ln(N)$ (with $[a_{i}]= [k]^{1}$ for  
$z=3$)~\cite{BPS}. This gives rise to a proliferation of independent
coupling constants. For example, for the minimal value of the dynamical 
critical exponent $z=3$, the number of independent terms in the
gravitational action of the non-projectable extension turns out to be
more than 70~\cite{KP}. In some regime of parameters, the
non-projectable extension is claimed to be free from the breakdown of
the standard perturbative expansion method in the $\lambda\to 1$ limit,
while in other regime the expansion breaks down. It is certainly
worthwhile performing a nonperturbative analysis of the non-projectable
theory in the regime of parameters where the standard perturbative
expansion breaks down and then identifying the observationally viable
regime of parameters.

There is yet another extension of the HL theory, with an additional
local $U(1)$ symmetry, $U(1) \ltimes {\mbox{Diff}}(M, \; {\cal{F}})$  
\cite{HMT}. It has been shown that the standard perturbative expansion
does not break down in the gravitational sector, but does break down in
the matter sector, at least apparently \cite{HW} (see also
\cite{Silva,WWa,Kluson,ZWWS,LWWZ} for more on this extension). It is intriguing
to see if a nonperturbative analysis similar to those presented in the
present paper can resolve this problem.

Finally, the biggest obstacle in front of the HL gravity, and in
general, any Lorentz symmetry breaking theory, is the restoration of the
Lorentz invariance in the matter sector at low energies
\cite{Collins:2004bp,Iengo:2009ix}. Even if the Lorentz violation is restricted only
to the gravity sector, the radiative corrections from graviton loops
will generate Lorentz violation in the matter sector. Such terms can be
under control provided that the Lorentz breaking scale is much lower 
than the Planck scale \cite{Pospelov:2010mp}.  Another approach is to
introduce a mechanism, or symmetry to suppress the Lorentz violating
operators at low energies, such as supersymmetry \cite{SUSY}. 
Such an approach was adopted in \cite{Xue:2010ih} where a SUSY theory with anisotropic scaling was constructed. On the other hand,  this seems to be a highly nontrivial task for the case of interacting models \cite{Redigolo:2011bv,Pujolas:2011sk}.

\begin{acknowledgments} 
 Part of this work was done during AW's visit at  IPMU, the University
 of Tokyo, Kashiwa. He would like to express his gratitude to the
 Institute for stimulating atmosphere and warm hospitality. The work of
 A.E.G. and S.M. was supported by the World Premier International
 Research Center Initiative (WPI Initiative), MEXT, Japan. S.M. also
 acknowledges the support by Grant-in-Aid for Scientific Research
 17740134, 19GS0219, 21111006, 21540278, by Japan-Russia Research
 Cooperative Program. AW is supported in part by DOE  Grant,
 DE-FG02-10ER41692. 

\end{acknowledgments}

\appendix

\section{Order ${\cal O}(\epsilon^{n+1})$ momentum constraint for $n\geq 2$}
\label{proof}

In this Appendix, we prove by induction that the order ${\cal O}(\epsilon^n)$ solution (\ref{eq:nthK})--(\ref{eq:nthgamma}) satisfies the order ${\cal O}(\epsilon^{n+1})$ momentum constraint equation (\ref{eq:n-const}) for $n\geq 2$. 

The proof extends the method presented in Ref.\cite{Izumi:2011eh} to include a scalar field source: we rewrite the left hand side of the 
($n+1$)-th order constraint (\ref{eq:n-const}) as a linear combination
of lower order constraints by using the explicit solution
(\ref{eq:nthK})--(\ref{eq:nthgamma}). To achieve this, we make use of the generalized Bianchi identity (\ref{eq:nth-DZ}) as well as the identities in Eq.(\ref{eqn:nth-identities}). We also use the following identity for
functions $f(t)$ and $g(t)$ satisfying
$a^3(t_{\rm in})\,f(t_{\rm in})\,g(t_{\rm in})=0$, 
\begin{equation}
 f(t)\,g(t) = \frac{1}{a^3(t)}\int_{t_{\rm in}}^t dt'\,a^3(t')
  \left[a(t')^{-3}\partial_{t'}(a^3(t')\,f(t'))\cdot g(t')
   + f(t')\cdot \partial_{t'}g(t')\right]. \label{eqn:integration-by-part}
\end{equation}

By applying the identity (\ref{eqn:integration-by-part}) to
$(f(t),g(t))=(A^{(p)\, j}_{\ \ \ \ \ i},\partial_j\zeta^{(n-p)})$,~ $(f(t),g(t))=(A^{(p)\, j}_{\ \ \ \ \ l},\, (\gamma^{-1})^{(q)\, lk}\partial_{i}\gamma^{(n-p-q)}_{jk})$ and $(f(t),g(t))=(\partial_t\phi^{(p)},\partial_i\phi^{(n-p)})$, 
the left hand side of the ($n+1$)-th order momentum constraint equation
(\ref{eq:n-const}) is rewritten as
\begin{eqnarray}
 {\cal C}^{(n+1)}_i & \equiv & 
  \partial_j A^{(n)\, j}_{\ \ \ \ \ i}
  +3 \sum_{p=1}^n A^{(p)\, j}_{\ \ \ \ \ i}\partial_j\zeta^{(n-p)}
  - \frac{1}{2}\sum_{p=1}^n\sum_{q=0}^{n-p}
  A^{(p)\, j}_{\ \ \ \ \ l}(\gamma^{-1})^{(q)\, lk}
  \partial_{i}\gamma^{(n-p-q)}_{jk} \nonumber\\
 & & 
  -\frac{1}{3}(3\,\lambda-1)\partial_i
  \left(K^{(n)}+\frac{3\,\phi^{(n)}\partial_t\phi^{(0)}}{3\,\lambda-1}
  \right)
  - \sum_{p=1}^{n-1}\partial_t\phi^{(p)}\partial_i\phi^{(n-p)}
  \nonumber\\
  &=& 
   \partial_j A^{(n)\, j}_{\ \ \ \ \ i}
   + \frac{1}{a^3(t)}\int_{t_{\rm in}}^t dt' a^3(t')
   \left\{
    3\, \sum_{p=1}^n\left[
      a^{-3}\partial_{t'}
      \left( a^3A^{(p)\, j}_{\ \ \ \ \ i}\right)
      \partial_j\zeta^{(n-p)}
      + A^{(p)\, j}_{\ \ \ \ \ i}
      \partial_j\left(\partial_{t'}\zeta^{(n-p)}\right)
     \right]
    \right. \nonumber\\
 & & 
  - \frac{1}{2}\sum_{p=1}^n\sum_{q=0}^{n-p}
  \left[ a^{-3}\partial_{t'}
   \left( a^3A^{(p)\, j}_{\ \ \ \ \ l}\right)
   (\gamma^{-1})^{(q)\, lk}
   \partial_{i}\gamma^{(n-p-q)}_{jk}
   + A^{(p)\, j}_{\ \ \ \ \ l}
   \partial_{t'}\left((\gamma^{-1})^{(q)\, lk}\right)
  \partial_{i}\gamma^{(n-p-q)}_{jk}\right.
  \nonumber\\
 & & \left.\left.
   + A^{(p)\, j}_{\ \ \ \ \ l}
   (\gamma^{-1})^{(q)\, lk}
  \partial_{i}\left(\partial_{t'}\gamma^{(n-p-q)}_{jk}\right)
  \right]
 -\sum_{p=1}^{n-1}
 \left[a^{-3}\partial_{t'}(a^3\partial_{t'}\phi^{(p)})
  \partial_i\phi^{(n-p)}
  + \partial_{t'}\phi^{(p)}
  \partial_{t'}\partial_i\phi^{(n-p)}\right]
     \right\}\nonumber\\
 & & 
  -\frac{1}{3}(3\,\lambda-1)\partial_i
  \left(K^{(n)}
   +\frac{3\,\phi^{(n)}\partial_t\phi^{(0)}}{3\,\lambda-1}\right)\,. 
\end{eqnarray}
Using Eqs.(\ref{eq:nthK})--(\ref{eq:nthA}), (\ref{eq:n-phi})--(\ref{eq:n-gamma})
and (\ref{eqn:gamma-inv-nth-eq}), this is further rewritten as 
\begin{eqnarray}
 {\cal C}^{(n+1)}_i 
  &=& 
  \frac{1}{a^3(t)}\int_{t_{\rm in}}^t dt' a^3(t')
  \left\{
   \partial_j
   \left( -\sum_{p=1}^{n-1}K^{(p)}A^{(n-p)\, j}_{\qquad\ \ i}
   \right)
    \right. \nonumber\\
 & & 
   +3\left[
      \sum_{p=2}^n
      \left(-\sum_{q=1}^{p-1}K^{(q)}A^{(p-q)\, j}_{\qquad\ \ i}\right)
      \partial_j\zeta^{(n-p)}
      + \sum_{p=1}^{n-1}A^{(p)\, j}_{\ \ \ \ \ i}
      \partial_j\left(\frac{1}{3}K^{(n-p)}\right)
	    \right]
   \nonumber\\
 & & 
  - \frac{1}{2}\sum_{p=1}^n\sum_{q=0}^{n-p}
  \left[ 
   \left(-\sum_{r=1}^{p-1}K^{(r)}A^{(p-r)\, j}_{\qquad\ \ l}\right)
   (\gamma^{-1})^{(q)\, lk}
   \partial_{i}\gamma^{(n-p-q)}_{jk}
   \right.\nonumber\\
 & &
  \left.
   + A^{(p)\, j}_{\ \ \ \ \ l}
   \left(-2\sum_{r=1}^q
  A^{(r)\, l}_{\quad\ \ m}
  (\gamma^{-1})^{(q-r)\, mk}\right)
  \partial_{i}\gamma^{(n-p-q)}_{jk}\right.
  \nonumber\\
 & & \left.
   + A^{(p)\, j}_{\ \ \ \ \ l}
   (\gamma^{-1})^{(q)\, lk}
   \partial_{i}
   \left(
    2 \sum_{r=0}^{n-p-q-1}\gamma^{(r)}_{jm}
    A^{(n-p-q-r)\, m}_{\qquad\qquad\quad k}
	   \right)\right]
  \nonumber\\
 & & 
 -\frac{1}{6}(3\,\lambda-1)\partial_i 
 \left(
 -\sum_{p=1}^{n-1} K^{(p)}K^{(n-p)} 
 \right)
 -\frac{1}{2}\partial_i
 \left(
  -\sum_{p=1}^{n-1}A^{(p)\, j}_{\ \ \ \ \, k}
  A^{(n-p)\, k}_{\qquad\ \ j}
       \right)
  \nonumber\\
 & & \left.
 + \sum_{p=1}^{n-1}\sum_{q=1}^p
  K^{(q)}\partial_{t'}\phi^{(p-q)}\partial_i\phi^{(n-p)}
 + \frac{1}{6}\sum_{p=1}^n\bar{Z}^{(p)}
 \sum_{q=0}^{n-p}
  (\gamma^{-1})^{(q)\, jk}
  \partial_{i}\gamma^{(n-p-q)}_{jk}
     \right\}\,,
\end{eqnarray}
where we have used the generalized Bianchi identity
(\ref{eq:nth-DZ}). By using the identities (\ref{eqn:nth-identities}) we
finally obtain 
\begin{equation}
 {\cal C}^{(n+1)}_i = 
  -\frac{1}{a^3(t)}\int_{t_{\rm in}}^t dt' a^3(t')
  \sum_{p=1}^{n-1}K^{(n-p)}{\cal C}^{(p+1)}_i. 
\end{equation}
Since the ${\cal O}(\epsilon^2)$ constraint in Eq.~(\ref{const-1}) is already satisfied, i.e. ${\cal C}^{(2)}_i=0$, the above relation implies that ${\cal C}^{(n+1)}=0$ for $n\geq 2$.

\section{Expansion of the nonlinear perturbations}
\label{app:expansion}

Here, we present the detail of the calculations to obtain the
expression of $\zeta$ in terms of $\zeta_T$ and $h_{ij}$, given in
Eq.(\ref{zetaNL}). While the former $\zeta$ is defined in the
synchronous gauge $N_i=0$, the latter $\zeta_T$ and $h_{ij}$ are defined
in in the transverse gauge $\delta^{ik} \partial_k h_{ij}=0$. In both
gauges, the freedom in the time coordinate is fixed by the choice
$N=1$. For the perturbative expansion of the spatial metric, we use 
\begin{equation}
g_{ij} = a^2 \,e^{2\,\zeta}\,\left( e^{h}\right)_{ij} = a^2 \delta_{ij} + \Bigg(a^2\left(2\,\zeta \delta_{ij} + h_{ij}\right) \Bigg) +\Bigg[\frac{a^2}{2}\,\left(4\,\zeta^2\delta_{ij}+4\,\zeta h_{ij}+h_{il}h^l_{\;j}\right)\Bigg]+{\cal O}(\bar{\epsilon}^3)\,,
\end{equation}
where $\bar{\epsilon}$ denotes the order of perturbations and the
indices of $h_{ij}$ are raised and lowered with Kronecker
delta. Throughout this Appendix, when the expansion of a quantity is
shown, the terms outside parentheses, in parentheses and in square
brackets are of order $\bar{\epsilon}^0$, $\bar{\epsilon}^1$ and
$\bar{\epsilon}^2$, respectively.

\subsection{Expansion of the momentum constraint}

We first concentrate on the linear perturbation in the transverse
gauge. To remove the nondynamical degrees in the shift vector, we solve
the constraint equation order by order. We expand the shift vector while 
separating the contributions from each order, as 
\begin{equation}
N_i = \Bigg(\partial_i \,B^{(1)} + N_i^{T\,
(1)}\Bigg)+\Bigg[\frac{1}{2}\left(\partial_i \,B^{(2)} + N_i^{T\,
(2)}\right)\Bigg]+{\cal O}(\bar{\epsilon}^3)
\end{equation}
With these decompositions, we expand the momentum constraint in vacuum 
\begin{equation}
{\cal H}_j\equiv D_iK^i_{\ j}- \lambda \,\partial_jK = 0 \,,
\end{equation}
as a series in perturbations. At first order, we get
\begin{equation}
{\cal H}^{(1)}_j=-\left(3\,\lambda-1\right)\partial_j\partial_t\zeta_T + \frac{\lambda-1}{a^2}\,\partial_j \triangle B^{(1)} - \frac{1}{2\,a^2}\triangle N^{T\,(1)}_j\,,
\end{equation}
which can be solved by
\begin{equation}
\triangle B^{(1)} = \frac{3\,\lambda - 1}{\lambda -1} a^2\,\partial_t\zeta_T \,,\qquad N_i^{T\,(1)} = 0\,.
\label{conslin}
\end{equation}
Using the second of these results, the next order constraint yields, 
\begin{eqnarray}
{\cal H}^{(2)}_j &=&\left(\lambda-1\right)\, \partial_j \left[\frac{1}{2}\,\triangle B^{(2)} + (\partial^k \zeta_T) \,(\partial_k B^{(1)}) -2\,\zeta_T \, \triangle B^{(1)} -h^{kl} \partial_k\partial_l B^{(1)}\right]\nonumber\\
&& +\frac{a^2}{4} \left[ (\partial_t h^{kl}) (\partial_k h_{lj}) - h^{kl} \partial_k \partial_t h_{lj} - (\partial_t h^{kl}) (\partial_j h_{kl})\right] + \frac{3\,a^2}{2}\,(\partial^k\zeta_T) \,(\partial_t h_{kj})\nonumber\\
&&+ (\partial_k\partial_j \zeta_T)\,(\partial^k B^{(1)}) + (\triangle \zeta_T) \, (\partial_jB^{(1)}) + \frac{1}{2}\,(\triangle h_{kj})(\partial^k B^{(1)}) - \frac{1}{4}\, \triangle N_j^{T\,(2)} = 0\,.
\label{consqua}
\end{eqnarray}
For the following, only the longitudinal part of this relation is
needed; by taking its divergence, then using the first equation of
(\ref{conslin}), we end up with 
\begin{eqnarray}
\triangle B^{(2)} &=& 2\,a^2 \left(\frac{3\,\lambda-1}{\lambda-1}\right) \left[2\,\zeta_T\,\partial_t\zeta_T + h^{ij}\partial_i\partial_j \triangle^{-1}\partial_t\zeta_T - (\partial^i \zeta_T) \,(\partial_i \triangle^{-1} \partial_t\zeta_T)\right] \nonumber\\
&&- \frac{2\,a^2\,\left(3\,\lambda-1\right)}{\left(\lambda-1\right)^2}\,\triangle^{-1}\,\left[2\,(\partial^i \triangle\zeta_T)(\partial_i \triangle^{-1}\partial_t\zeta_T) + \left(\partial^i\partial^j \zeta_T+ \frac{1}{2}\,\triangle h^{ij}\right) (\partial_i \partial_j \triangle^{-1}\partial_t\zeta_T)    + (\triangle\zeta_T)\,(\partial_t \zeta_T) \right]\nonumber\\
&&+ \frac{a^2}{\lambda-1}\,\triangle^{-1}\,\left[ \frac{1}{2}\,(\partial_i \partial_t h_{jk})\,(\partial^i h^{jk}) + \frac{1}{2}\,(\partial_t h_{ij})\, (\triangle h^{ij}) -3 \,(\partial_i\partial_j\zeta_T)\,(\partial_t h^{ij})\right]\,.
\label{consqualon}
\end{eqnarray}

\subsection{Coordinate transformations}

Next, we determine the transformation between the transverse and
synchronous gauges. We parametrize the coordinate transformation as 
\begin{equation}
\tilde{x}^\mu  = x^\mu + \left(\xi^{(1)\,\mu}\right) + \left[\frac{1}{2} (\xi^{(1)\,\nu} \partial_\nu \xi^{(1)\,\mu}+  \xi^{(2)\,\mu})\right] + {\cal O}(\bar{\epsilon}^3)\,,
\label{transseries}
\end{equation}
where over-tilde denotes quantities in the synchronous gauge. The
parameters $\xi^{(n)\,\mu}$ are decomposed as 
\begin{equation}
\xi^{(n)\,\mu} = \left(0,\, \xi^{(n)\,i}+\partial^i \xi^{(n)}\right)\,,
\end{equation}
with $\partial_i\xi^{(n)\,^i}=0$, while the indices of $\xi^{(n)\,i}$
are raised and lowered by $\delta^{ij}$ and $\delta_{ij}$. For any
tensor field expanded as
\begin{equation}
T = T^{(0)} + \Big(\delta T\Big)+ \left[\frac{1}{2}\, \delta^2 T\right] + {\cal O}(\bar{\epsilon}^3)\,,
\end{equation}
the transformation at linear and quadratic order proceeds through
\cite{Bruni:1996im} 
\begin{eqnarray}
\widetilde{\delta T} &=& \delta T + \pounds_{\xi^{(1)}} T^{(0)} \,,\nonumber\\
\widetilde{\delta^2 T} &=& \delta^2 T + 2\,\pounds_{\xi^{(1)}} \delta T +\pounds_{\xi^{(1)}}^2 T^{(0)}+  \pounds_{\xi^{(2)}} T^{(0)} \,.
\end{eqnarray}
For the metric tensor, the transformations become
\begin{eqnarray}
\widetilde{\delta g_{\mu\nu}} &=& \delta g_{\mu\nu} + g_{\mu\sigma}^{(0)} \partial_\nu \xi^{(1)\,\sigma} + g_{\nu\sigma}^{(0)}\partial_{\mu}\xi^{(1)\,\sigma}\,,\nonumber\\
\widetilde{\delta^2 g_{\mu\nu}} &=& \delta^2 g_{\mu\nu} + 2\,\left(\xi^{(1)\,\sigma}\,\partial_\sigma \delta g_{\mu\nu} + \delta g_{\mu\sigma} \partial_\nu \xi^{(1)\,\sigma}+ \delta g_{\nu\sigma} \partial_\mu \xi^{(1)\,\sigma}\right)\nonumber\\
&& + g_{\mu\sigma}^{(0)}\,\partial_\nu\left(\xi^{(1)\,\rho} \partial_\rho \xi^{(1)\,\sigma}\right) + g_{\nu\sigma}^{(0)}\,\partial_\mu\left(\xi^{(1)\,\rho} \partial_\rho \xi^{(1)\,\sigma}\right) + 2\,g_{\rho\sigma}^{(0)} (\partial_\mu \xi^{(1)\,\sigma})(\partial_\nu \xi^{(1)\,\rho})\nonumber\\
&& + g_{\mu\sigma}^{(0)} \partial_\nu \xi^{(2)\,\sigma} + g_{\nu\sigma}^{(0)}\partial_{\mu}\xi^{(2)\,\sigma}\,.
\label{gmunu2}
\end{eqnarray}
We now determine the transformation $\xi^\mu$ needed to go from the
transverse gauge to the synchronous gauge. For this, we evaluate the
$0i$ components of (\ref{gmunu2}) and set 
$\widetilde{\delta g_{0i}}=0$. At first order, we obtain, 
\begin{equation}
\partial_i B^{(1)} + N_i^{T\,(1)} + a^2\,\left(\partial_i \partial_t\xi^{(1)} + \partial_t \xi_i^{(1)}\right)=0\,,
\end{equation}
where the transverse and longitudinal parts can be easily separated to give,
\begin{equation}
\xi^{(1)} = - \int^t dt' \,\frac{B^{(1)}(t')}{a^2}\,,
\qquad
\xi^{(1)}_i = - \int^t dt'\,\frac{N_i^{T\,(1)}(t')}{a^2}\,.
\end{equation}
Using the solutions (\ref{conslin}) to the linear momentum constraint,
the transformation parameters become 
\begin{equation}
\xi^{(1)} = -\frac{3\,\lambda -1}{\lambda-1}\,\triangle^{-1}\,\zeta_T\,,
\qquad
\xi^{(1)}_i = 0\,.
\label{xilin}
\end{equation}
Similarly, the $0i$ component of the second order transformation
(\ref{gmunu2}) gives,  
\begin{equation}
\partial_i B^{(2)} + N_i^{T\,(2)} + a^2 \left(\partial_i \partial_t \xi^{(2)}+\partial_t\xi_i^{(2)}\right)+ (\partial^j\xi^{(1)})\,(\partial_i\partial_j B^{(1)})-(\partial_i\partial^j\xi^{(1)})\,(\partial_j B^{(1)})
-4\,\zeta_T\,\partial_i B^{(1)} -2\,h_{ij}\partial^j B^{(1)}=0\,,
\end{equation}
where we used the second equation of (\ref{xilin}). Using also the first
equation of (\ref{xilin}) as well as the first order constraint
(\ref{conslin}), the longitudinal part of the second order
transformation can be obtained as 
\begin{eqnarray}
\triangle\xi^{(2)} &=& -\int^t dt' \,\frac{\triangle B^{(2)}}{a^2} + \left(\frac{3\,\lambda-1}{\lambda-1}\right)^2 \left(\partial^i\zeta_T\right)\,\left(\partial_i\triangle^{-1} \zeta_T\right)+2\,\left(\frac{3\,\lambda-1}{\lambda-1}\right)\,\zeta_T^2\nonumber\\
&& +2\,\left(\frac{3\,\lambda-1}{\lambda-1}\right)\int^t dt'\,\left[h^{ij}\partial_i \partial_j \triangle^{-1}\partial_{t'}\zeta_T-\left(\frac{\lambda+1}{\lambda-1}\right)(\partial^i\zeta_T)\left(\partial_i\triangle^{-1}\partial_{t'}\zeta_T\right)\right]\,.
\end{eqnarray}
Inserting the expression of $B^{(2)}$ from (\ref{consqualon}) into the
above expression, we finally obtain 
\begin{eqnarray}
\triangle\xi^{(2)} &=& \left(\frac{3\,\lambda-1}{\lambda-1}\right)^2 \left(\partial^i\zeta_T\right)\,\left(\partial_i\triangle^{-1} \zeta_T\right)-\frac{4\,(3\,\lambda-1)}{(\lambda-1)^2}\,\int^t dt'\,(\partial_i\zeta_T)\,(\partial^i \triangle^{-1}\partial_{t'}\zeta_T)
\nonumber\\
&&+ \frac{2\,\left(3\,\lambda-1\right)}{\left(\lambda-1\right)^2}\int^t dt'\triangle^{-1}\left[2\,(\partial^i \triangle\zeta_T)(\partial_i \triangle^{-1}\partial_{t'}\zeta_T) + \left(\partial^i\partial^j \zeta_T+ \frac{1}{2}\,\triangle h^{ij}\right) (\partial_i \partial_j \triangle^{-1}\partial_{t'}\zeta_T)    + (\triangle\zeta_T)(\partial_{t'} \zeta_T) \right]\nonumber\\
&&-\frac{1}{\lambda-1}\int^t dt'\,\triangle^{-1}\left[ \frac{1}{2}\,(\partial_i \partial_{t'} h_{jk})(\partial^i h_{jk}) + \frac{1}{2}\,(\partial_{t'} h^{ij})(\triangle h_{ij}) -3\,(\partial_i\partial_j\zeta_T)(\partial_{t'} h^{ij})\right]
\,.
\label{xiqua}
\end{eqnarray}
We note that the second order gauge transformation is more divergent
than the first order one (\ref{xilin}), in the limit $\lambda\to1$.

Finally, we calculate the field $\zeta$ in the synchronous gauge. Since
we adopted a nonperturbative decomposition for the spatial metric, it is
useful to express this quantity as, 
\begin{equation}
\zeta = \frac{1}{6}\,\log\left(\frac{\det\,\tilde{g}}{a^6}\right).
\end{equation}
Applying the perturbative expansion to the right hand side, we obtain
\begin{equation}
\zeta= \left(\frac{1}{6\,a^2}\,\widetilde{\delta g_{ii}}\right) 
+\left[\frac{1}{12\,a^2}\,\left(\widetilde{\delta^2 g_{ii}} -\frac{1}{a^2}\widetilde{\delta g_{ij}}\,\widetilde{\delta g_{ij}}\right)
\right]
+{\cal O}(\bar{\epsilon}^3)\,.
\label{zetsync}
\end{equation}
Using the transformed metric from (\ref{gmunu2}), the above expression
becomes 
\begin{equation}
\zeta = \left(\zeta_T +\frac{1}{3}\,\triangle \xi^{(1)}\right) + \left[(\partial^i\zeta_T)\,(\partial_i \xi^{(1)}) + \frac{1}{6} (\partial^i\xi^{(1)})\,(\partial_i \triangle\xi^{(1)}) + \frac{1}{6}\,\triangle \xi^{(2)}\right]+{\cal O}(\bar{\epsilon}^3)\,.
\end{equation}
Finally, using the transformations (\ref{xilin}) and (\ref{xiqua}), we
obtain Eq.(\ref{zetaNL}), which relates the nonlinear perturbation
$\zeta$ in the synchronous gauge to an expansion series of perturbations 
$\zeta_T$ and $h_{ij}$ in the transverse gauge.

\end{document}